\documentclass[twocolumn,aps,prl,showpacs,amsmath,superscriptaddress,longbibliography,notitlepage]{revtex4-2}
\usepackage{amssymb}
\usepackage{mathrsfs}
\usepackage{graphicx}
\usepackage{float}
\usepackage[caption=false]{subfig}
\usepackage{xcolor}
\usepackage[colorlinks,linkcolor={red!75!black},citecolor={red!75!black},urlcolor={blue!75!black}]{hyperref}
\usepackage{verbatim}
\usepackage{epstopdf}
\usepackage[normalem]{ulem}
\usepackage{xcolor}
\usepackage{bm}
\usepackage{soul}
\usepackage{ulem}
\begin{document}
\title{Impurity-induced topological decomposition}
\author{Tianxing Shi}\thanks{These authors contribute equally to this work.}
\affiliation{Guangdong Provincial Key Laboratory of Quantum Metrology and Sensing $\&$ School of Physics and Astronomy, Sun Yat-sen University (Zhuhai Campus), Zhuhai 519082, China}
\author{Chuhang Zhang}\thanks{These authors contribute equally to this work.}
\affiliation{School of Physics, Nankai University, Tianjin 300071, China}
\author{Liang Jin}\email{jinliang@nankai.edu.cn}
\affiliation{School of Physics, Nankai University, Tianjin 300071, China}
\author{Linhu Li}\email{lilinhu@quantumsc.cn}
\affiliation{Quantum Science Center of Guangdong-Hong Kong-Macao Greater Bay Area (Guangdong), Shenzhen, China}

\begin{abstract}
Controlling topological phases is a central goal in quantum materials and related fields, enabling applications such as robust transport and programmable edge states.  Here we uncover a mechanism in which local on-site impurities act as knobs to decompose global topological properties in discrete steps. In non-Hermitian lattices with spectral winding topology, we show that each impurity sequentially reduces the winding number by one, which is directly manifested as a stepwise decomposition of quantized plateaus in the steady-state response. Based on this principle, we further develop a scheme that sequentially induces topological edge states under impurity control, in a class of Hermitian topological systems constructed by doubling the non-Hermitian ones. Our findings reveal a general scheme to tune global topological properties with local disruptions, establishing a universal framework for impurity-controlled topological phases and offering a foundation for future exploration of reconfigurable topological phenomena across diverse physical platforms.
\end{abstract}

\maketitle

{\it Introduction}.---
%Topological phases of matter exhibit remarkable phenomena, such as robust edge states and quantized responses, protected by global topological structures of energy bands~\cite{hasan2010colloquium,qi2011topological,RevModPhys.88.021004,RevModPhys.88.035005,xie2021higher}.
Topological phases of matter exhibit remarkable phenomena such as edge localization and quantized responses, arising from the global topological structures of energy bands.
In Hermitian systems, band topology is generally characterized by integer invariants defined under periodic boundary conditions (PBCs), which predict the number of edge states protected by a band gap under open boundary conditions (OBCs)~\cite{hasan2010colloquium,qi2011topological,RevModPhys.88.021004,RevModPhys.88.035005,xie2021higher}.
Recent advances in non-Hermitian physics have further uncovered a unique class of spectral winding topology
\cite{gong2018topological,okuma2020topological,zhang2020correspondence,wang2021generating,zhang2021acoustic,su2021direct,cao2023probing,wang2024non,yang2025observing,liang2025topological}, manifested through the non-Hermitian skin effect~\cite{yao2018edge,yao2018non,martinez2018non,jin2019bulk,lee2019anatomy,okuma2023non,ding2022non,zhang2022review,lin2023topological} and quantized response plateaus in the steady-state response governed by non-Hermitian Hamiltonian~\cite{li2021quantized,liu2021exact,liang2022anomalous,ou2023non}.
In this rapidly developing field, 
%a central challenge is not only to identify distinct topological phases but also to control and modulate them in a systematic manner, which is essential for both fundamental understanding and potential applications.
a central task for both fundamental understanding and potential applications is not only to identify distinct topological phases but also to control and modulate them in a systematic manner.
Typically, this can be implemented through topological phase transitions, which usually require changing global parameters of a system.

\begin{figure}
		\includegraphics[width=1\linewidth]{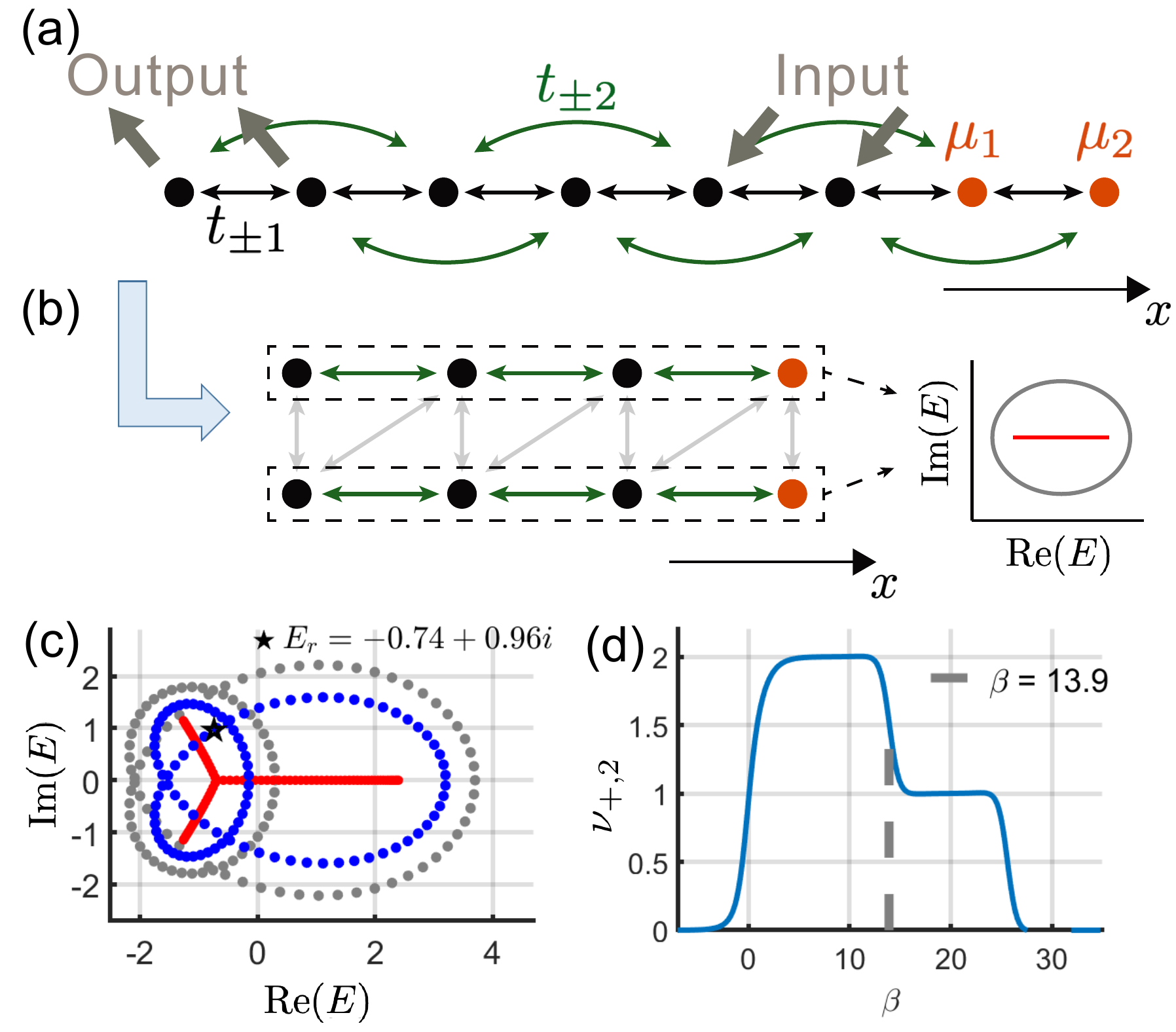}  
		\caption{(a) The model of Eq. \eqref{nonH_H} with nearest ($t_{\pm1}$ and next-nearest ($t_{\pm2}$) hopping and $N_{\mu}=2$. $\mu_1$ and $\mu_2$ represent the two impurities. $\nu_{+,2}$ defined in Eq. \eqref{eq:response} 
measures the growth rate of the response at the two left-hand sites (the "output") as the impurity strength increases when an external field is added to the two right-hand sites (the "input").
(b) %An intuitive understanding of the decomposition of spectral winding topology.
The model in (a) can be rearranged into a multi-chain lattice with only nearest-neighbor hopping along $x$ direction. Omitting the inter-chain coupling ($t_j$ with $|j|<{\rm max}[r,l]$), each chain corresponds to a loop-like spectrum under PBCs with winding number $|W(E_r)|=1$  (gray loop on the right), which can be trivialized into a line-spectrum by a single strong impurity that effectively induces OBCs (red line on the right).
The overall spectral winding (summing over all chains) is thus reduced by $1$ for each impurity.
(c) Energy spectra of the model under PBCs without (gray) and with two identical impurities ($\mu_1=\mu_2=e^{13.9}$, blue), and under OBCs (red). 
(d) Response quantity $\nu_{+,2}$ as a function of $\beta$, for a reference energy $E_r=-0.74+0.96i$ [black star in (b)]. 
The plateau at $\nu_{+,2}=2$ drops to $1$ when $\beta\approx 13.9$, where the spectrum in (b) goes through the reference energy $E_r$.
Other parameters are $t_{1}=1$, $t_{-1}=0.7$, $t_{2}=2$,$t_j=0$ for other values of $j$, and $N=50$.
}
		\label{fig1}  
	\end{figure}

%Conventionally, different topological properties can be tuned through topological phase transitions, which are implemented through changing global parameters.
%In contrast, here 
In this paper, we uncover a mechanism by which local impurities act as precise knobs to decompose and thereby control the global topological features in discrete steps.
Specifically, in one-dimensional (1D) non-Hermitian lattices with arbitrary spectral winding topology generated by long-range coupling, each on-site impurity reduces the spectral winding number by one, and further produces a stepwise decomposition of quantized response plateaus.
Physically, such a topological decomposition can be understood by rearranging the lattice into a series of 1D chains with only nearest-neighbour hopping, each possessing a spectral winding number of $1$, as sketched in Figs. \ref{fig1}(a) and \ref{fig1}(b).
When these chains are decoupled, a single impurity trivializes the spectral winding topology only for the chain it resides, thus reducing the overall spectral topology number by one.
Intriguingly, we find that such a topological decomposition remains robust even when different chains are coupled together, representing a universal topological feature insensitive to system parameters.
Remarkably, the formalism can be further extended to 
construct Hermitian topological phases and manipulate the topological edge states therein through local impurities, without involving topological phase transitions that require changing global parameters. 
Namely, topological edge states can be generated sequentially as impurities are added, which directly links the decomposition of winding topology to Hermitian boundary phenomena.
These findings establish a general strategy for selectively manipulating global topological properties through local disruptions,
%by which local perturbations can selectively manipulate global topological properties,
offering a practical route to control edge states and quantized responses in quantum devices and simulations.
%offering a practical route for reconfigurable topological phases in quantum devices and simulations.

{\it Non-Hermitian spectral winding and quantized steady-state response}.---
%\section{model and quantized steady-state response}
We first consider a single-band non-Hermitian model with long-range hopping and local on-site impurity potentials [as sketched in Fig. \ref{fig1}(a)], described by the Hamiltonian
\begin{eqnarray}
&&{H}={H}_0+{H}_\mu,\label{nonH_H}\\
{H}_0=\sum_{x=1}^N\sum_{j=-r}^l t_j \hat{c}^\dagger_x&&\hat{c}_{x+j},~~
{H}_{\mu}=\sum_{x=1}^{N_\mu}\mu_x\hat{c}^\dagger_{N-x+1}\hat{c}_{N-x+1},\nonumber
\end{eqnarray}
with $N$ the total number of lattice sites, $r$ ($l$) the maximal hopping range toward right (left) direction, $\hat{c}^\dagger_x$ creates a particle at site $x$,
and $t_j$ the hopping amplitude for the particle from sites $x+j$ to $x$.
The local impurity distributes on $N_\mu={\rm max}[r,l]\ll N$  sites near $x=N$, and $\mu_x$ is the impurity strength on site $N-x+1$.
In addition, $\hat{c}^\dagger_{x+N}=\hat{c}^\dagger_x$ is set to have ${H}_0$ describing a translational invariant system under the PBCs, while strong impurities isolated the $N_\mu$ lattice sites, effectively inducing OBCs to the rest of the system.

In general, this model supports a loop-like PBC spectrum when any $t_j\neq t_{-j}^*$, characterized by a spectral winding number~\cite{gong2018topological,okuma2020topological,zhang2020correspondence}
\begin{eqnarray}\label{eq:winding}
W(E_r)=\oint_{-\pi}^{\pi}\frac{dk}{2\pi}\frac{d}{dk}{\rm arg}[H(k)-E_r],
\end{eqnarray}
with $H(k)=\sum_{j=-r}^l t_j e^{ikj}$ the Bloch Hamiltonian of ${H}_0$ and $E_r$ a reference energy for defining the winding.
By increasing the strength of a uniform impurity $\mu_x=e^\beta$, the system gradually evolves from PBCs to OBCs,
during which the winding number manifests as a quantized quantity in the steady-state response governed by the non-Hermitian Hamiltonian~\cite{li2021quantized}, as shown in Fig. \ref{fig1}(c) and (d) for an example with $N_\mu=2$ and $|W(E_r)|\leqslant2$.
Explicitly, the quantized response quantity is defined as~\cite{li2021quantized}
\begin{align}\label{eq:response}
\nu_{\alpha,m}=d\ln |G_{\alpha,m\times m}|/d\beta,
\end{align}
with $m=|W(E_r)|$, $\alpha=\pm$ the sign of $W(E_r)$, and $G_{+,m\times m}$ ($G_{-,m\times m}$) the $m\times m$ matrix block at the top-right (bottom-left) corners (negleting the impurity sites) of the Green's function $G=1/(E_r-{H})$.
%(see Supplemental Materials for more details \LLH{[May reproduce some calculations from the NC paper in SupMat]}). 
It is seen in Fig. \ref{fig1}(c) that when increasing $\beta$, the loop-like spectrum under PBCs gradually evolves to the Y-shape spectrum under OBCs;
and the response quantity in Fig. \ref{fig1}(d) forms quantized plateaus that drops from $2$ to $1$ as the spectrum evolves through the chosen reference energy $E_r$.
The other drop from $1$ to $0$ corresponds to the second passage of the spectrum through $E_r$ and  is not shown in the figure for simplicity.
%passing through the chosen reference energy $E_r$ a number of times equal to $|W(E_r)|$.
%During this process, the response quantity jumps between plateaus quantized at the spectral winding number, as shown in Fig.\ref{fig1}(c).

\begin{figure*}
		\includegraphics[width=1\linewidth]{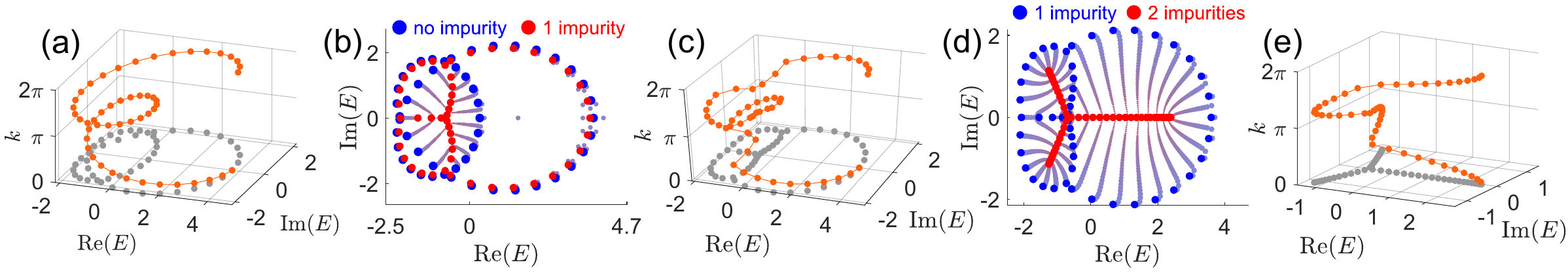}  
		\caption{Spectral winding and evolution with increasing impurity strengths.
		(a) Spectral winding for our model under PBCs in the absence of impurities, $\mu_1=\mu_2=0$.
		Gray dots are the spectrum on the complex energy plane, and orange dots and line show its trajectory with quasi-momentum $k$ varying from $0$ to $2\pi$.
		(b) Spectral evolution (lilac color) with $\mu_2=0$ and $\mu_1$ increasing from $0$ (blue) to $e^{18}$ (red). Further increasing $\mu_1$ does not significantly affect the spectrum in the absence of the second impurity.
		(c) The same as in (a) but with $\mu_1=e^{18}$, where the spectral winding number is reduced to $W(E_r)=1$ for the area enclosed by the complex spectrum. %\red{Different colors (orange, blue, and green) indicate the three regions with $W(E_r)=1$.}
		(d) The same as in (b), but with $\mu_1=10^{35}$ and $\mu_2$ increasing from $0$ to $e^{35}$. 
		(e) The same as in (a), but with $\mu_1=\mu_2=10^{35}$. The spectrum no longer encloses any area and $W(E_r)=0$ for arbitrary $E_r$, as under OBCs.
		Other parameters are $t_{1}=1$, $t_{-1}=0.7$, $t_{2}=2$,$t_j=0$ for other values of $j$, and $N=50$.
		Impurity states with exponentially large eigenenergies are omitted in all panels.
		}  
		\label{fig2}  
	\end{figure*}

{\it Decomposition of non-Hermitian spectral topology}.---
We now describe a more intriguing scenario with non-uniform impurity strengths on different sites,
which decomposes a large spectral winding number $|W(E_r)|>1$ into smaller ones.
Specifically, we add the two impurity potential $\mu_1$ and $\mu_2$ separately to the previous system with $W(E_r)=2$.
As can be seen in Figs. \ref{fig2}(a) and \ref{fig2}(b), with increasing a single impurity $\mu_1$, the PBC spectrum gradually evolves into its interior, but cannot reach the Y-shape OBC spectrum in Fig. \ref{fig1}(c), as the two ends of the system are still connected by nonzero $t_{\pm 2}$ even with the single impurity site $x=N$ removed from the lattice.
Instead, when $\mu_1$ exceeds a critical value,
only the region with $W(E_r)=2$ shrinks and vanishes,
and the eigenenergies form three adjacent loops that share parts of their boundaries.
Note that the spectral winding number is ill-defined for the impurity system due to the lack of the Bloch momentum $k$.
Nevertheless, we may still assign the $k$-index to each eigenenergy, by following the continuous spectral evolution from PBCs to $\mu_1\rightarrow \infty$ in Fig. \ref{fig2}(b).
A winding direction can thus be determined for the impurity spectrum in Fig. \ref{fig2}(c), which has $W(E_r)=1$ in the interiors of the loops. 
Finally, by further introducing and increasing a second impurity $\mu_2$ on site $x=N-1$, the adjacent-loop spectrum continues to shrinks into the Y-shape OBC spectrum without nontrivial spectral winding, as shown in Figs. \ref{fig2}(d) and \ref{fig2}(e).
In other words, the spectral topology with a large winding number $W(E_r)=2$ is decomposed and reduced to $W(E_r)=1$ by a single on-site impurity ($\mu_1$), and trivialized by further introducing a second one ($\mu_2$).

\begin{figure}
		\includegraphics[width=1\linewidth]{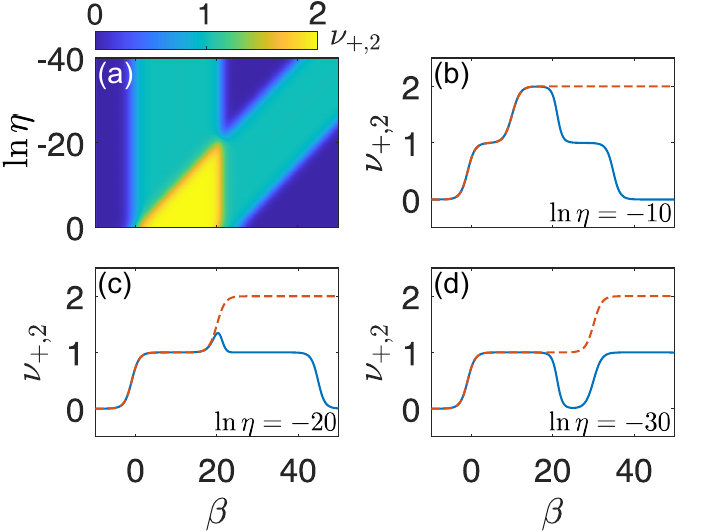}  
		\caption{Quantized response quantity $\nu_{+,2}$ for impurities $\mu_2={\eta}\mu_1={\eta}e^{\beta}$ and reference energy $E_r=-1+0.15i$.
		(a) $\nu_{+,2}$ varying with $\beta$ for different ratio $\eta$ between the two impurities.
		The plateau at $\nu_{+,2}=2$ is decomposed to two at $\nu_{+,1}$ when $|\ln \eta|$ increases.
		(b) to (d) Numerical (blue solid lines) and analytical (red dash lines) results of the quantized response at $\ln \eta=-10, -20, -30$, respectively. The drops of the numerical plateaus correspond to the saturation of response amplification due to finite system size, thus they do not occur in analytical results obtained under thermodynamic limit.
		Other parameters are the same as in Fig. \ref{fig2}.
		}  
		\label{fig3}  
	\end{figure}

The decomposition of spectral winding topology can be further reflected in the quantized steady-state response.
To see this, we consider the response quantity $\nu_{+,2}$ with fixed relative ratio $\eta$ between the amplitudes of the two impurities,
$$\mu_2={\eta}\mu_1={\eta}e^{\beta}.$$ 
%so that the system can be tuned between the single- and double-impurity scenarios.
%Thus, the system recovers the identical-impurity scenario in Fig. \ref{fig1} when ${\eta}=0$. 
For ${\eta}\ll1$, the system is expected to first enter the single-impurity scenario in Figs. \ref{fig2}(b) and \ref{fig2}(c) for a relatively small $\beta$, 
then the double-impurity scenario in Figs. \ref{fig2}(d) and \ref{fig2}(e) when $\beta$ is large enough.
As can be seen in Fig. \ref{fig3}(a), the plateau of $\nu_{+,2}=2$ at $\eta=1$ is gradually decomposed into two at $\nu_{+,2}=1$ when decreasing the ratio $\eta$.
Specifically for our chosen parameters, the quantized response is partially decomposed with $\ln \eta\gtrsim -20$, 
where the plateau of $\nu_{+,2}=2$ survives in a certain range of $\beta$ [see Fig. \ref{fig3}(b)].
%where the spectral evolutions in Figs. \ref{fig2}(b) and \ref{fig2}(d) occurs simultaneously.
On the other hand, a complete decomposition occurs at a larger magnitude of $\ln \eta$, as indicated by the two fully separated plateaus of $\nu_{+,2}=1$. In particular,
the first (second) plateau with smaller (larger) $\beta$ corresponds to the spectral evolution in Fig. \ref{fig2}(b) [(d)], and drops to zero when eigenenergies pass through the reference energy $E_r$ during the evolution.

The quantized response plateaus and their decomposition can also be predicted analytically.
First, for PBC Hamiltonian $H_0$ without any impurity, the elements of its Green's function $G_0$ can be solved with Bloch wave function, 
given by
\begin{align}
[G_0]_{xy}\sum_n(E_r-t_j\sum_{j=-r}^{l}e^{i j k_n})^{-1}\frac{e^{ik_n(x-y)}}{N}
\end{align}
with $k_n=2n\pi/N$, $n=1,2,3,...N$.
Next, adding a single impurity $\mu_{x_0}$ at site $x_0$,
the Green's function $G^{x_0}$ of the new Hamiltonian can be obtained as~\cite{economou2006green,leonforte2021vacancy,lombardo2014photon,roccati2021non}
\begin{align}\label{eq:G_impurity}
G^{x_0}=G_0+\mu_{x_0}\frac{G_0|x_0\rangle\langle x_0|G_0}{1-\mu_{x_0}\langle x_0|G_0|x_0\rangle}.
\end{align}
Repeating this calculation, we can in principle solve the Green's function for the system with an arbitrary number $N_\mu$ of on-site impurities, yet the computational complexity increases drastically with $N_\mu$.
Nevertheless, for $N_\mu=2$, the response quantity in the thermodynamic limit can be obtained as~\cite{SuppMat}
\begin{align}
\nu^{\rm ana}_{+,2}=
 \frac{e^{\beta}[2\eta e^{\beta}-E_r(1+\eta)]}{-t_1t_{-1}+(E_r-e^{\beta})(E_r-\eta e^{\beta})}.
\end{align}

In Figs. \ref{fig3}(b) to \ref{fig3}(d), it is seen that $\nu^{\rm ana}_{+,2}$ accurately predict the increase of $\nu_{+,2}$, but not its decrease.
To understand this divergence, note that $\nu_{+,2}$ corresponds to the growth rate of the response amplitude (with respect to $\beta$), which increases exponentially with the system's size $N$ under OBCs~\cite{wanjura2020topological,xue2021simple,wanjura2021correspondence}.
In a finite-size system, strengtherning impurities eventually isolate the lattice sites they reside and induce OBCs to the rest of the system at finite $\beta$ values, corresponding to the saturation of the response amplitude and the drop of the plateau of $\nu_{+,2}$.
However, these finite-size feature cannot be captured by our analytical solution obtained under the thermodynamic limit, where the response amplitude tends to infinity under OBCs.
%The drop of a plateau of $\nu_{+,2}$ indicates the saturation of the response amplitude associating to an impurity of the finite-size system, thus it cannot be captured by the analytical solution obtained under the thermodynamic limit.

Finally, we note that even in a finite system, the decomposition of the high response plateau generally does not coincide with the decomposition of spectral winding, as the former also relies on the chosen reference energy $E_r$.
For an intermedia value of ${\eta}$, the spectral evolutions in Figs. \ref{fig2}(b) and \ref{fig2}(d) may simultaneously occur in a small parameter region of $\beta$, meaning that the spectral winding topology is only partially decomposed;
yet the quantized response can still be decomposed completely into plateaus at $\nu_{+,2}=1$, provided the response amplitude associated with the
stronger impurity ($\mu_1$ in our examples) already saturates for the choose $E_r$ before the system entering this parameter region (see Supplemental Materials for an example \cite{SuppMat}).

{\it Decomposition of Hermitian band topology}.---
To further generalize the scope of topological decomposition,
we now extend the scheme to develop a formalism for controlling Hermitian topological properties through local impurities in a similar manner.
%The decomposition of spectral winding number in non-Hermitian systems further provides a formalism to decompose Hermitian topological properties in a similar manner.
Explicitly, we construct a doubled Hermitian Hamiltonian ${H}_{\rm H}$ with ${H}$ in Eq. \eqref{nonH_H}~\cite{gong2018topological,lee2019topological},
\begin{align}
{H}_{\rm H}=
\begin{pmatrix}
0&H-E_r\\
H^\dagger-E_r^*&0
\end{pmatrix},\label{eq:H_H}
\end{align}
which possesses a pseudospin-1/2 degree of freedom through the doubling.
It satisfies a chiral symmetry $\sigma_zH_{\rm H}\sigma_z=-H_{\rm H}$ that protects topological properties in 1D~\cite{schnyder2008classification,ryu2010topological},
with $\sigma_z$ a Pauli matrix acting at the doubled pseudospin space. 
In particular, the winding number $W(E_r)$ defined in Eq. \eqref{eq:winding} predicts the number of pairs of topological edge states under OBCs.

\begin{figure}
		\includegraphics[width=1\linewidth]{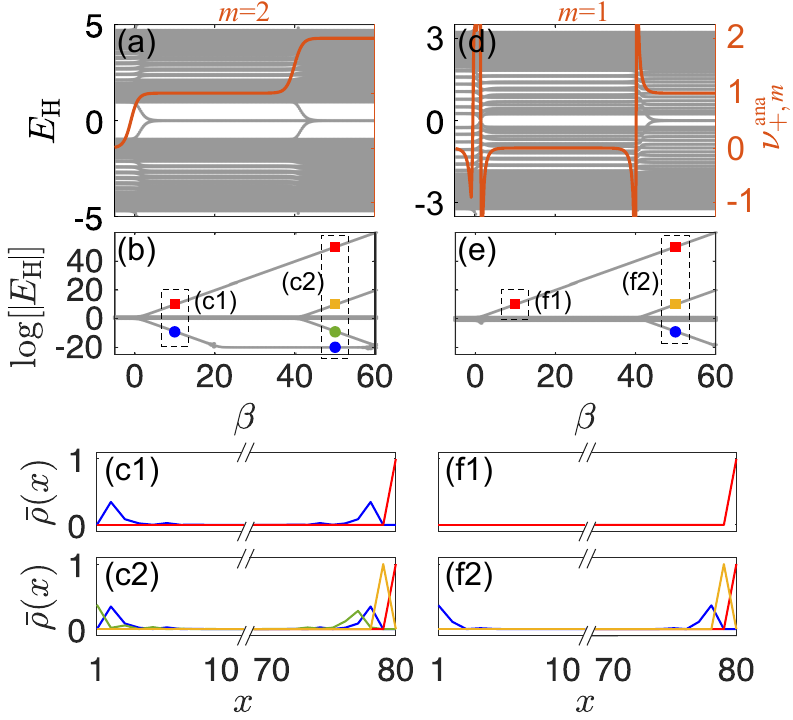}  
		\caption{Topological properties of the the doubled Hermitian Hamiltonian $H_{\rm H}$.
		(a) OBC spectrum of $H_{\rm H}$ with $E_r=-1+0.15i$ and $W(E_r)=2$, in comparison of the quantized response quantity $\nu^{\rm ana}_{+,2}$ for the non-Hermitian Hamiltonian $H$.
		A pair of zero-energy edge states appear when $\nu^{\rm ana}_{+,2}$ increases by $1$ at $\beta\approx 0$ and $40$.
		(b) Logarithm of the spectrum in (a), where impurity states and topological edge states feature exponentially increasing and decreasing eigenenergies, respectively.
		Note that these states are both two-fold degenerate in ${\rm log}[|E_{\rm}|]$.
		(c) Distribution of impurity and topological edge states marked by the same colors (squares or circles) in (b).
		Note that each mark corresponds to a pair of eigenstates;
		and $\bar{\rho}(x)$ represents the average density of them summed over the two pseudospin components in Eq.~\eqref{eq:H_H}.
		(d) to (f) the same as (a) to (c), but with $E_r=0.5+0.15i$ and $W(E_r)=1$. 
		Impurity states appear similarly at the two critical values of $\beta$,
		while topological edge states and nonzero plateau of $\nu_{+,1}$ appear only at the latter ($\beta\approx 40$).
		Other parameters are $t_1=1$, $t_{-1}=0.7$, $t_2=2$, $t_{-2}=0$, $\eta=e^{-40}$, and $N=80$.
}  
		\label{fig4}  
\end{figure}

In Figs. \ref{fig4}(a) and \ref{fig4}(b), we demonstrate the spectrum $E_{\rm H}$ of ${H}_{\rm H}$ for $W(E_r)=2$ with strongly detuned impurities ($\eta=e^{40}$),  in comparison to the analytical result of quantized response quantity, $\nu^{\rm ana}_{+,2}$, of the non-Hermitian Hamiltonian $H$.
%Specifically, in Fig. \ref{fig3}(a), we consider the case with $W(E_r)=2$ under PBCs, and strongly detuned impurities with $c=e^{40}$ that fully decompose the topological response into two plateaus at $\nu_{2,+}=1$.
By gradually increasing $\beta$,
the impurities $\mu_1$ and $\mu_2$ 
induce impurity states when $\beta$ approximately exceed $0$ and $40$, respectively, whose eigenenergies exponentially growing with $\beta$ [red and yellow squares in (b)].
At each of these critical values of $\beta$,
the quantized response quantity $\nu^{\rm ana}_{+,2}$ increases by $1$, and a pair of zero-energy edge states [blue and green circles in (b)] emerge from the bulk bands of $H_{\rm H}$.
Distributions of these impurity and zero-energy edge states are illustrated in Fig. \ref{fig4}(c).
The two pairs of zero-energy states emerged at different $\beta$ constitute the $2W(E_r)=4$ topological edge states of $H_{\rm H}$ when $\beta\rightarrow \infty$, 
where both impurity sites are isolated and the rest of the system is effectively under OBCs.
In other words, 
each impurity induces one pair of topological zero-energy states to $H_{\rm H}$, representing a decomposition of the Hermitian topology with $W(E_r)=2$.
Note that in contrast to non-Hermitian systems, the Hermitian topological properties are generally size-independent.
Thus the zero-energy edge states do not capture the drops of plateaus in finite-size non-Hermitian systems (see Fig. \ref{fig3}) either.

Finally in Figs. \ref{fig4}(d) to \ref{fig4}(f), we display another example with $W(E_r)=1$ as a comparison, where $H_{\rm H}$ supports only one pair of topological edge states under OBCs.
Similarly to the previous case, impurity states with exponentially growing eigenenergies emerge also at both $\beta\approx0$ and $40$, regardless of the different topological nature of the system.
However, 
a nonzero plateau with $\nu_{+,1}=1$ and a pair of zero-energy edge states appear only when $\beta\gtrsim40$, where both impurities are strong enough to isolate the lattice sites from the rest of the system.
Finally, in addition to the example discussed here, we also provide another case with
$W(E_r) = 3$ in Supplemental Materials~\cite{SuppMat} as a further verification of the discovered mechanism.

%\section{Multi-band models}
%\LLH{[This is a less important section, which can be moved to SuppMat, or simply removed]}

{\it Conclusion and discussion}.---
We have shown that local impurities can act as precise control knobs to decompose topological properties in discrete steps. In non-Hermitian lattices with spectral winding topology, each impurity reduces the winding number by one and induces a corresponding decomposition of quantized response plateaus. 
Such a topological decomposition can be comprehended by omitting short-range coupling and leaving only the longest-range one, resulting in a series of decoupled chains with $|W(E_r)|=1$ for each of them [see Fig. \ref{fig1}(b)].
The spectral winding topology for the overall system is simply described by the sum of winding numbers over all chains, which is reduced by one for each (strong) impurity that trivializes the spectral winding for the chain it resides.
More importantly, our results reveal that this decomposing mechanism remains robust even in the presence of inter-chain coupling (short-range coupling of the original model), verifying its universality as a hitherto unreported class of topological phenomena.
The same principle carries over to Hermitian systems through a doubled construction, where edge states emerge sequentially under impurity control.
These findings establish a general scheme for manipulating global topology with minimal local disruptions. 
Note that in practice, the model with exponentially strong impurity magnitudes can be obtained by simply removing the corresponding lattice sites.
Thus, beyond its conceptual interest, 
this impurity-induced topological decomposition may also facilitate controlling topological properties, such as edge states and quantized responses, in various quantum and classical platforms that realize rich topological phases.

In the current formalism, decomposition of Hermitian topology applies for the doubled Hamiltonian $H_{\rm H}$ with a chiral symmetry,
which protects a $Z$-type topology characterized by integer invariants.
In 1D, nontrivial topology can also be protected by the particle-hole symmetry, yet it is characterized by a $Z_2$ topology and does not process large topological invariant that can be decomposed.
However, in 2D and higher dimension, $Z$-type topology may exist even in the absence of chiral symmetry,
whose possible topological decomposition cannot be directly connected to non-Hermitian spectral winding with the formalism established here and requires further exploration.
%On the other hand, the correspondence between non-Hermitian systems and doubled Hermitian ones has also been extended to the study of hybrid skin-topological effect (or, higher-order non-Hermitian skin effect) and Hermitian higher-order topological phases~\cite{lee2019hybrid,kawabata2020higher,okugawa2020second,fu2021non,zhu2022hybrid,zhu2024brief}, which may provide a relatively simple generalization of our formalism into higher dimensions.

\begin{center}
\textbf{Acknowledgement}
\end{center}
This work is supported by 
the National Natural Science Foundation of China (Grant No. 12474159 and No. 12525502).

\clearpage

\onecolumngrid

\begin{center}
\textbf{\large Supplementary Materials}
\end{center}

\tableofcontents
\setcounter{secnumdepth}{3}

		%%%%%%%%%% Prefix a "S" to all equations, figures, tables and reset the counter %%%%%%%%%%
	\setcounter{equation}{0} \setcounter{figure}{0} \setcounter{table}{0} %
	\renewcommand{\theequation}{S\arabic{equation}} \renewcommand{\thefigure}{S%
		\arabic{figure}} 
	%\renewcommand{\citenumfont}[1]{S#1}
	%%%%%%%%%% Prefix a "S" to all equations, figures, tables and reset the counter %%%%%%%%%%

\section{Derivation of the quantized response quantity}
In this supplemental section we provide an analytical approach to the quantized response quantity in the steady-state response for our system with impurities, described by the Hamiltonian
\begin{eqnarray}
{H}&=&{H}_0+{H}_\mu,\\
{H}_0&=&\sum_{x=1}^N\sum_{j=-r}^l t_j \hat{c}^\dagger_x\hat{c}_{x+j},{H}_{\mu}=\sum_{x=1}^{N_\mu}\mu_x\hat{c}^\dagger_{N-x+1}\hat{c}_{N-x+1}.\nonumber
\end{eqnarray}
The Green's function of this system, regarding the complex reference energy $E_r$, is given by
\begin{eqnarray}
G_0=(E_r-H_0)^{-1}=\sum_n(E_r-E_n)^{-1}|\psi_{nR}\rangle\langle\psi_{nL}|,
\end{eqnarray}
with $|\psi_{nR}\rangle$ ($|\psi_{nL}\rangle$) the right (left) eigenvectors of the Hamiltonian matrix~\cite{kunst2018biorthogonal,brody2013biorthogonal}.
Under PBCs, the left and right eigenvectors of $H_0$ are the same, given by the Bloch state
\begin{eqnarray}
|\psi_n\rangle =\frac{1}{\sqrt{N}}\sum_{x} e^{ik_n x}\hat{c}^\dagger_x|0\rangle,
\end{eqnarray}
with $k_n= 2n\pi/N$, $n=1,2,...N$. Then the elements of the Green's function are given by
\begin{eqnarray}
[G_0]_{xy}=\sum_n(E_r-t_j \sum_{j=-r}^le^{ijk_n})^{-1}\frac{e^{ik_n(x-y)}}{N}.\label{eq:G0_sum}
\end{eqnarray}
Rewriting the summation into an integral, we obtain
\begin{eqnarray}
[G_0]_{xy}=\oint_{|z|=1}\frac{dz}{2\pi i}\frac{z^{x-y-1}}{E_r-\sum_j t_jz^j},\label{eq:G0_int}
\end{eqnarray}
with $z=e^{ik}$. Note that in the summation $k$ takes discrete value $k_n$, and we always have $Nk_n=2n\pi$. Therefore we can always take $z^N=1$ in the integral, which will be useful in the following calculation.

\subsection{A single impurity with a spectral winding number $|w(E_r)|=1$}
To begin with, we first consider a single impurity potential adding to the $N$th site. The total Hamiltonian is given by $H=H_0+H_\beta$, with
\begin{eqnarray}
H_\beta=\mu_1\hat{c}^\dagger_{N}\hat{c}_{N},
\end{eqnarray}
with $\mu_1=e^\beta$ the strength of the impurity.
The Green's function of the whole system reads~\cite{roccati2021non}
\begin{eqnarray}
G(\beta)=(E_r-H)^{-1}=G_0+e^{\beta}\frac{G_0 |N\rangle\langle N|G_0}{1-e^{\beta} \langle N|G_0|N\rangle}\label{eq:Green_beta}
\end{eqnarray}
with $|N\rangle=\hat{c}^\dagger_N|0\rangle$.

Now we consider each element of $G^{\beta}$. We have
\begin{eqnarray}
[G(\beta)]_{xy}&=&[G_0]_{xy}+e^{\beta}\frac{[G_0]_{xN}[G_0]_{Ny}}{1-e^{\beta}[G_0]_{NN}}\\
&=&\oint_{|z|=1}\frac{dz}{2\pi i}\frac{z^{x-y-1}}{E_r-\sum_j t_jz^j}+e^{\beta}\frac{
\oint_{|z|=1}\frac{dz}{2\pi i}\frac{z^{x-N-1}}{E_r-\sum_j t_jz^j}
\oint_{|z|=1}\frac{dz}{2\pi i}\frac{z^{N-y-1}}{E_r-\sum_j t_jz^j}
}{1-e^{\beta}\oint_{|z|=1}\frac{dz}{2\pi i}\frac{z^{-1}}{E_r-\sum_j t_jz^j}}.\nonumber\\
\end{eqnarray}
Next we shall consider the element connecting the two ends of the system, corresponding to the signal amplification across the 1D chain. As site $N$ is dominated by the impurity when $e^{\beta}$ is large, we shall only focus on the rest $N-1$ sites. For the element connecting sites $1$ and $N-1$, we have  (note that $z^N=1$)
\begin{eqnarray}
[G(\beta)]_{1(N-1)}&=&\oint_{|z|=1}\frac{dz}{2\pi i}\frac{z}{E_r-\sum_j t_jz^j}+e^{\beta}\frac{
\oint_{|z|=1}\frac{dz}{2\pi i}\frac{1}{E_r-\sum_j t_jz^j}
\oint_{|z|=1}\frac{dz}{2\pi i}\frac{1}{E_r-\sum_j t_jz^j}
}{1-e^{\beta}\oint_{|z|=1}\frac{dz}{2\pi i}\frac{z^{-1}}{E_r-\sum_j t_jz^j}}.\nonumber\\
\end{eqnarray}
In this equation we have three different integrals, which can be solved through the residue theorem. But before analysing the residues, we first take them as some constants $C_{1,2,3}$, and the equation becomes
\begin{eqnarray}
[G(\beta)]_{1(N-1)}=C_1+\frac{e^{\beta}C_2^2}{1-e^{\beta} C_3}:=C_1+f(\beta).\label{eq:C123}
\end{eqnarray}
Thus we have
\begin{eqnarray}
\nu_{+,1}:=\frac{\partial \ln [G(\beta)]_{1(N-1)}}{\partial \beta}&=&\frac{1}{C_1+f(\beta)}\frac{\partial f(\beta)}{\partial \beta}\nonumber\\
&=&\frac{f(\beta)}{C_1+f(\beta)}\frac{\partial \ln f(\beta)}{\partial \beta}\nonumber\\
&=&\frac{f(\beta)}{C_1+f(\beta)}\frac{1}{1-e^\beta C_3}.
\end{eqnarray}
Thus we can see that when $\beta$ gets larger, $\nu_{+,1}$ is non-vanishing only when $C_3=0$. Furthermore, we can see that $\nu_{+,1}$ is nearly quantized at $1$ provided $|f(\beta)|\gg |C_1|$ (namely $|[G(\beta)]_{1(N-1)}|\gg|C_1|=const.$) is also satisfied, meaning that there is a large amplification ratio for a signal entering from site $N-1$ and leaving at site $1$.

%These integrals are given by the sum of their residues $r_0$ and/or $r_m$, at $z_0=0$ and/or $z_m$ satisfying $E_r-\sum_j t_jz_m^j=0$, possibly with more than one $z_m$, provided $|z_0|<1$ and/or $|z_m|<1$. 
For the simplest case with only nonzero $t_{\pm1}$, namely the Hatano-Nelson model,
the denominator (and its zeros) of the integrals is given by
\begin{eqnarray}
E_r-t_1 z-t_{-1}/z&=&0\nonumber\\
t_1 z^2-E_r z+ t_{-1}&=&0,\nonumber\\
z_{1,2}=\frac{E_r\pm \sqrt{E_r^2-4t_1t_{-1}}}{2t_1}.
\end{eqnarray}

{\bf Case (i)} If $|z_{1,2}|<1$, we have
\begin{eqnarray}
C_1&=&\oint_{|z|=1}\frac{dz}{2\pi i}\frac{z^2}{-t_1(z-z_1)(z-z_2)}\nonumber\\
&=&\frac{1}{-t_1}(\frac{z_1^2}{z_1-z_2}+\frac{z_2^2}{z_2-z_1})\nonumber\\
&=&\frac{1}{-t_1}(z_1+z_2)=-\frac{E_r}{t_1^2},
\end{eqnarray}
\begin{eqnarray}
C_2&=&\oint_{|z|=1}\frac{dz}{2\pi i}\frac{z}{-t_1(z-z_1)(z-z_2)}\nonumber\\
&=&\frac{1}{-t_1}(\frac{z_1}{z_1-z_2}+\frac{z_2}{z_2-z_1})\nonumber\\
&=&-\frac{1}{t_1},
\end{eqnarray}
\begin{eqnarray}
C_3&=&\oint_{|z|=1}\frac{dz}{2\pi i}\frac{1}{-t_1(z-z_1)(z-z_2)}\nonumber\\
&=&\frac{1}{-t_1}(\frac{1}{z_1-z_2}+\frac{1}{z_2-z_1})\nonumber\\
&=&0,
\end{eqnarray}
and
\begin{eqnarray}
\nu_{+,1}&=&
\frac{e^\beta C_2^2}{C_1+e^\beta (C_2^2-C_1C_3)}\frac{1}{1-e^\beta C_3}\nonumber\\
&=&\frac{e^\beta C_2^2}{C_1+e^\beta C_2^2}=\frac{e^\beta }{-E_r+e^\beta},\label{eq:nu_i}
\end{eqnarray}
thus we have  $\nu_{+,1}=1$ when $e^\beta\gg |E_r|$ (i.e. $|f(\beta)|\approx|e^{\beta}C_2^2|\gg |C_1|$).

Consistently, the winding number of Eq. (2) in the main text is~\cite{zhang2020correspondence}
\begin{eqnarray}
W(E_r)=\oint_{-\pi}^{\pi}\frac{dk}{2\pi}\frac{d}{dk}{\rm arg}[H(k)-E_r]=N_{\rm zeros}-N_{\rm poles}=1,\label{eq:winding_ana}
\end{eqnarray}
where $N_{\rm zeros}$ ($N_{\rm poles}$) is the counting of zeros (poles) 
of the integral 
enclosed by the BZ ($|z|=1$) weighted by respective orders. Here we have two zeros at $z=z_{1,2}$ and one pole at $z=0$ with $N_{\rm poles}=1$.

\bigskip

{\bf Case (ii)} If $|z_{1,2}|>1$, we have $|z_1z_2|=|t_{-1}/t_1|>1$, the winding number $w(E_r)=-1$, and $C_{1,2,3}=0$. 
In such cases we shall obtain $[G(\beta)]_{1(N-1)}=0$, and $\nu_{+,1}$ is ill-defined as $\ln 0\rightarrow -\infty$. 
%Numerically, in a finite system, $G_{1(N-1)}^\beta$ shall take a small but nonzero value, and we shall $\nu_{1(N-1)}=0$.
On the other hand, we may exchange $t_1$ and $t_{-1}$, i.e. inverse the 1D chain (so that $N-1$ and $1$ are exchanged), and thus recover the scenario of case (i). Therefore we shall consider the other element $[G(\beta)]_{(N-1)1}$ and the quantity $\nu_{-,1}=\frac{\partial \ln [G(\beta)]_{(N-1)1}}{\partial \beta}$, which will be quantized at $1$.
%A simple picture is we can exchange $t_1$ and $t_{-1}$, i.e. inverse the 1D chain (so that $N-1$ and $1$ are exchanged), and we can have the same results as in case i).

\bigskip

{\bf Case (iii)} If $|z_1|<1$ and $|z_2|>1$, we have the winding number $w(E_r)=0$. For the coefficients $C_{1,2,3}$, we obtain
\begin{eqnarray}
C_1&=&-\frac{1}{E_r}z_1^2\nonumber\\
C_2&=&-\frac{1}{E_r}z_1=-\frac{E_r+\sqrt{E_r^2-4t_1t_{-1}}}{2E_rt_1},\nonumber\\
C_3&=&-\frac{1}{E_r}.
\end{eqnarray}
The defined quantity is given by
\begin{eqnarray}
\nu_{+,1}&=&
\frac{e^\beta C_2^2}{C_1+e^\beta (C_2^2-C_1C_3)}\frac{1}{1-e^\beta C_3}\nonumber\\
&=&\frac{e^\beta z_1^2}{-z_1^2 E_r}\frac{1}{1+e^\beta/E_r }\nonumber\\
&=&-\frac{e^\beta }{E_r+e^\beta},\label{eq:nu_iii}
\end{eqnarray}
thus we have  $\nu_{+,1}=-1$ when $e^\beta\gg |E_r|$. Similarly, we can also obtain $\nu_{-,1}=-1$.
%and we shall have $\nu_{1(N-1)}\approx0$ when $|e^\beta C_3|\gg1$, i.e. $e^\beta\gg |E_r|$.

{\bf Case (iv)} The last case with $|z_2|<1$ and $|z_1|>1$ is similar to this case, with $C_{1,2,3}$ having $z_2$ replacing $z_1$ and an extra minus sign.

\bigskip

From the above results we see that the value of $\nu_{1(N-1)}$, and also that of $\nu_{(N-1)1}$, is independent from the exact parameters of the system, but only depends on the reference energy $E_r$ and the impurity strength $e^\beta$. 
Especially, when the impurity is strong enough (i.e. $e^\beta\gg |E_r|$), these two quantities can take only $\pm1$, depending solely on the spectrum winding number of the system. Thus we have the following correspondence between the spectral winding number and the defined quantities:
\begin{eqnarray}
W(E_r)=1\rightarrow [\nu_{+,1},\nu_{-,1}]=[1,-];\nonumber\\
W(E_r)=-1\rightarrow [\nu_{+,1},\nu_{-,1}]=[-,1];\nonumber\\
W(E_r)=0\rightarrow [\nu_{+,1},\nu_{-,1}]=[-1,-1].\nonumber
\end{eqnarray}
Here the sign "$-$" means that the corresponding quantity is ill-defined [see case (ii)]. 
%On the other hand, the divergence of the element of the Green's function, say $[G(\beta)]_{1(N-1)}$ in case (ii), is obtained in the thermodynamic limit (as we have replaced the summation with an integral). In a finite system, $G_{1(N-1)}^\beta$ shall take a small but nonzero value as it describes the signal propagation in two sites near the impurity, which is also assume to be finite in numerical approach. So that numerically we shall obtain $\nu_{1(N-1)}=0$ in such cases. Thus the spectral winding number can be simply given by
%\begin{eqnarray}
%w(E_r)=\nu_{1(N-1)}-\nu_{(N-1)1}.
%\end{eqnarray}

\begin{figure}
		\includegraphics[width=1\linewidth]{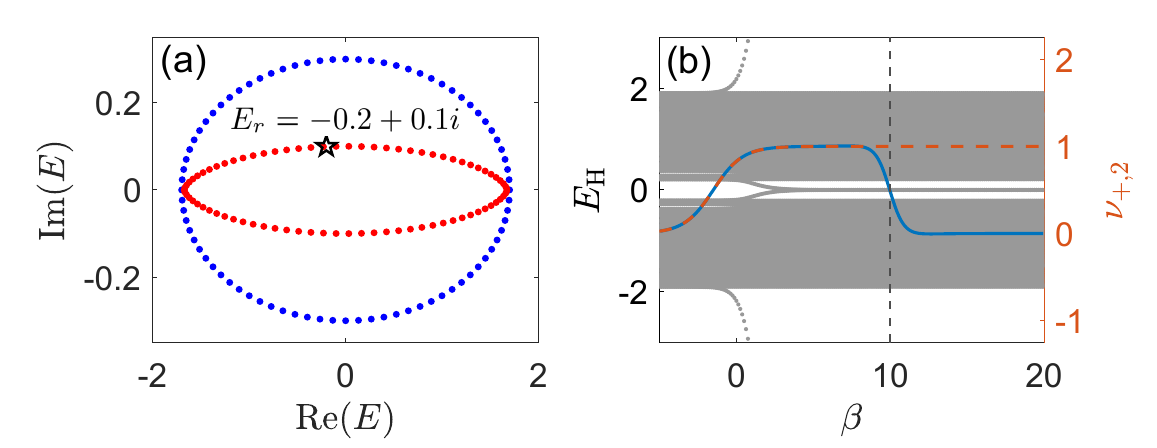}  
		\caption{(a) Complex eigenenergies for our model with only nearest-neighbor hopping. 
		The spectrum under PBCs (blue dots) possesses $W(E_r)=1$ for the chosen reference energy $E_r=-0.2+0.1i$ (black star),
		and passes through $E_r$ in the presence of an extra impurity $\mu_1=e^{\beta}$ with $\beta=10$ (red dots).
		(b) Quantized response quantity $\nu_{+,1}$ obtained numerically (blue solid line), and analytically (red dash line) from Eq. \eqref{eq:nu_i}. The plateau at $\nu_{+,1}=1$ ends at $\beta\approx 10$, where the spectrum pass through $E_r$ in (a).
		In addition, we display eigenenergies of the doubled  Hermitian Hamiltonian $H_{\rm H}$ (see Eq. (7) in the main text), where the emergence of zero-energy topological edge states concides with that of the quantized plateau.
		Other parameters are $t_1=1$, $t_{-1}=0.7$, $t_j=0$ for other values of $j$, and $N=80$.
		}\label{sub:fig1}  
\end{figure}

In Fig. \ref{sub:fig1}, we compare the numerical  and analytical results of $\nu_{+,1}$ for the nontrivial case with $W(E_r)=1$, which agree well with each other [and also with the emergence of zero-energy topological edge states in a double Hermitian Hamiltonian, see Eq.~(7) of the main text].
Note that numerically, $\nu_{+,1}$ jumps to zero when $\beta$ exceeds a certain value, which cannot be captured by our analytical results [Fig.~\ref{sub:fig1}(b)]. 
This is because in the calculation of Green's function elements, we have rewritten the summation into an integral [see Eq.~\eqref{eq:G0_int}],
meaning that the derivations assume the thermodynamic limit.
On the other hand, $[G_0]_{1(N-1)}$ describes the response amplitude at site $1$ for a driving field at site $N-1$;
in the absence of impurity, it vanishes under PBCs but grows exponentially with the system size $N$ under OBCs, $[G_0]_{1(N-1)}\sim \alpha^{N-1}$ with $\alpha$ being a real number (which corresponds to a directional amplification of the response, see Refs.~\cite{wanjura2020topological,xue2021simple,wanjura2021correspondence}).
In our numerical results, we consider a finte-size PBC system with an impurity $\mu_1=e^\beta$ added to the $N$th site.
The response quantity $\nu_{+,1}$ describes the growing rate of the $[G(\beta)]_{1(N-1)}$ with $\beta$, and its drop from $1$ to $0$ 
means that the response amplitude $[G(\beta)]_{1(N-1)}$ reaches its maximal value ($\sim \alpha^{N-1}$) for the chosen reference energy $E_r$ [after which the spectrum no longer encloses th $E_r$, see Fig. \ref{sub:fig1}(a)].
However, the response quantity $\nu_{+,1}$, which describes the growing rate of $[G(\beta)]_{1(N-1)}$ with $\beta$, is finite.
Thus, in the thermodynamic limit, $[G(\beta)]_{1(N-1)}$ can never reach the OBC limit with  $[G_0]_{1(N-1)}\rightarrow \infty$ at any finite $\beta$, so that our analytical results of $\nu_{+,1}$ never drop to $0$.

\subsection{Two impurities with a spectral winding number $|W(E_r)|=2$}
For more impurities, the Green's function can be solved by repeatedly using Eq.~\eqref{eq:Green_beta}, however, the computational complexity increases drastically with the number of impurities.
Here we extend our analysis to the case with two impurities added to sites $N$ and $N-1$, in an extended Hatano-Nelson model with next-nearest neighbor hopping.
The Hamiltonian is given by
\begin{eqnarray}
{H}&=&{H}_0+{H}_\mu,\nonumber\\
{H}_0&=&\sum_{x=1}^N\sum_{j=-2}^2 t_j \hat{c}^\dagger_x\hat{c}_{x+j},{H}_{\mu}=\sum_{x=1}^{2}\mu_x\hat{c}^\dagger_{N-x+1}\hat{c}_{N-x+1},
\end{eqnarray}
with $\mu_2=\eta\mu_1=\eta e^\beta$.
Introducing nonzero $t_{\pm2}$ can generate a spectral winding number up to $|W(E_r)|=2$,
and the corresponding response quantity is given by~\cite{li2021quantized} 
$\nu_{\pm,2}=d\ln |G_{\pm,2\times 2}|/d\beta$ with
\begin{eqnarray}
G_{+,2\times2}=\begin{pmatrix}
    [G(\beta)]_{1(N-3)}      & [G(\beta)]_{1(N-2)}  \\
    [G(\beta)]_{2(N-3)}       & [G(\beta)]_{2(N-2)}  
\end{pmatrix},~~
G_{-,2\times2}=\begin{pmatrix}
    [G(\beta)]_{(N-3)1}      & [G(\beta)]_{(N-2)1}  \\
    [G(\beta)]_{(N-3)2}       & [G(\beta)]_{(N-2)2}  
\end{pmatrix}.
\end{eqnarray}

Similar to the previous case, the Green's function element is determined by [see Eq.~\eqref{eq:Green_beta}]
\begin{eqnarray}
G(\beta)=G_1+\eta e^{\beta}\frac{G_1 |N\rangle\langle N|G_1}{1-\eta e^{\beta} \langle N|G_1|N\rangle},\\
G_1(\beta)=G_0+e^{\beta}\frac{G_0 |N\rangle\langle N|G_0}{1- e^{\beta} \langle N|G_0|N\rangle},\\
\end{eqnarray}
with $G_0$ the Green's function of the PBC system without any impurity.
As can be seen from Eq.~\eqref{eq:G0_int} (and the previous example with a single impurity),
elements of $G_0$ is determined by the zeros of $E_r-H_0$,
which is given by
\begin{align}
E_r-t_2 z^2-t_1 z-t_{-1}/z-t_{-2}/z^2=0
\end{align}
in the current case.
Here we sort its solutions by their modulus, $|z_1|\leq|z_2|\leq|z_3|\leq|z_4|$.
Without loss of generality, we consider two nontrivial cases with $W(E_r)=2$ and $W(E_r)=1$ in the following discussion, where we have omitted the variant $\beta$ in $G(\beta)$ for the sack of simplicity.

\subsubsection{$W(E_r)=2$}
Following Eq.~\eqref{eq:winding_ana}, the system has $W(E_r)=2$ when $|z_{1,2,3,4}|<1$ (with a single order-2 pole at $z=0$).
With some further derivations, we obtain
\begin{eqnarray*}
	%G_{-1}&=&\left[G\right]_{N-1,N}=0,\\
	%G_0&=&\left[G\right]_{N,N}=\left[G\right]_{N-1,N-1}=0,\\
	%G_1&=&\left[G\right]_{1,N}=\left[G\right]_{N,N-1}=0,\\
	G_2&=&\left[G\right]_{1,N-1}=\left[G\right]_{N-1,N-3}=\left[G\right]_{N,N-2}=\left[G\right]_{2,N}=-\frac{1}{t_2},\\
	G_3&=&\left[G\right]_{N,N-3}=\left[G\right]_{2,N-1}=\left[G\right]_{1,N-2}=\frac{t_1}{\left(t_2\right)^2},\\
	G_4&=&\left[G\right]_{1,N-3}=-\frac{\left(t_1\right)^2}{\left(t_2\right)^3}-\frac{E_r}{\left(t_2\right)^2},\\
	G_5&=&\left[G\right]_{2,N-3}=\frac{\left(t_1\right)^3}{\left(t_2\right)^4}+\frac{2E_rt_1}{\left(t_2\right)^3}+\frac{t_{-1}}{\left(t_2\right)^2},
\end{eqnarray*}	
and
\begin{eqnarray}
G_{+,2\times2}&=&\left[G_4+\eta e^{\beta}\left(G_2\right)^2\right]\left[G_4+e^{\beta}\left(G_2\right)^2\right]-G_3\left[G_5+e^{\beta}G_2G_3\left(1+\eta\right)\right],\\
	\nu_{+,2} &=& \frac{e^{\beta}[2\eta e^{\beta}-E_r(1+\eta)]}{-t_1t_{-1}+(E_r-e^{\beta})(E_r-\eta e^{\beta})}.
\end{eqnarray}
 The response quantity $\nu^{\rm ana}_{+,2}$ in Figs. 3 and 4(a) in the main text is obtained using the above equation.

\subsubsection{$W(E_r)=1$}
The system has $W(E_r)=1$ when $\left\vert z_4\right\vert>1$ and $\left\vert z_{1,2,3}\right\vert<1$. In this case, the quantized response quantity is given by $\nu_{+,1}=d\ln |[G(\beta)]_{1(N-2)}|/d\beta$.
It is difficult to obtain the explicit form of $\nu_{+,1}$ in this case;
nevertheless, the Green's function element can be obtained as
\begin{equation}
G_{1,N-2} =G_{3}+\mu_{\scriptscriptstyle{N}}e^{\beta }\frac{G_{1}G_{2}}{1-\mu_{\scriptscriptstyle{N}}e^{\beta }G_{0}}%
+\mu_{\scriptscriptstyle{N-1}}e^{\beta }\frac{\left( G_{2}+\mu_{\scriptscriptstyle{N}}e^{\beta }\frac{G_{1}G_{1}}{%
		1-\mu_{\scriptscriptstyle{N}}e^{\beta }G_{0}}\right) \left( G_{1}+\mu_{\scriptscriptstyle{N}}e^{\beta }\frac{G_{-1}G_{2}%
	}{1-\mu_{\scriptscriptstyle{N}}e^{\beta }G_{0}}\right) }{1-\mu_{\scriptscriptstyle{N-1}}e^{\beta }\left(
	G_{0}+\mu_{\scriptscriptstyle{N}}e^{\beta }\frac{G_{-1}G_{1}}{1-\mu_{\scriptscriptstyle{N}}e^{\beta }G_{0}}\right) },
\end{equation}
with
\begin{eqnarray*}
	G_{-1}&=&\left[G\right]_{N-1,N}=\frac{1}{t_2\left(z_4-z_1\right)\left(z_4-z_2\right)\left(z_4-z_3\right)},\\
	G_0&=&\left[G\right]_{N,N}=\left[G\right]_{N-1,N-1}=\frac{z_4}{t_2\left(z_4-z_1\right)\left(z_4-z_2\right)\left(z_4-z_3\right)},\\
	G_1&=&\left[G\right]_{1,N}=\left[G\right]_{N,N-1}=\frac{\left(z_4\right)^2}{t_2\left(z_4-z_1\right)\left(z_4-z_2\right)\left(z_4-z_3\right)},\\
	G_2&=&\left[G\right]_{1,N-1}=\left[G\right]_{N-1,N-3}=\left[G\right]_{N,N-2}=\left[G\right]_{2,N}=-\frac{1}{t_2}+\frac{\left(z_4\right)^3}{t_2\left(z_4-z_1\right)\left(z_4-z_2\right)\left(z_4-z_3\right)},\\
	G_3&=&\left[G\right]_{N,N-3}=\left[G\right]_{2,N-1}=\left[G\right]_{1,N-2}=\frac{t_1}{\left(t_2\right)^2}+\frac{\left(z_4\right)^4}{t_2\left(z_4-z_1\right)\left(z_4-z_2\right)\left(z_4-z_3\right)}.
	%G_4&=&\left[G\right]_{1,N-3}=-\frac{\left(t_1\right)^2}{\left(t_2\right)^3}-\frac{E_r}{\left(t_2\right)^2}+\frac{\left(z_4\right)^5}{t_2\left(z_4-z_1\right)\left(z_4-z_2\right)\left(z_4-z_3\right)}\\
	%G_5&=&\left[G\right]_{2,N-3}=\frac{\left(t_1\right)^3}{\left(t_2\right)^4}+\frac{2E_rt_1}{\left(t_2\right)^3}+\frac{t_{-1}}{\left(t_2\right)^2}+\frac{\left(z_4\right)^6}{t_2\left(z_4-z_1\right)\left(z_4-z_2\right)\left(z_4-z_3\right)}\\
\end{eqnarray*}
The response quantity $\nu^{\rm ana}_{+,1}=\partial \ln|G_{1,N-2}|/\partial \beta$ in Fig. 4(d) of the main text is obtained numerically from the above results.

\section{Deviation between decomposition of quantized response and spectral winding}
In the main text, we have unveiled the impurity-induced decomposition of both the overall winding topology of the non-Hermitian spectrum, and its corresponding quantized steady-state response that reflects the winding regarding a specific reference energy $E_r$.
Due to their independence/dependence on $E_r$, the decomposition of these topological features is expected to occurs at different impurity strengths.

\begin{figure}[H]
		\includegraphics[width=0.9\linewidth]{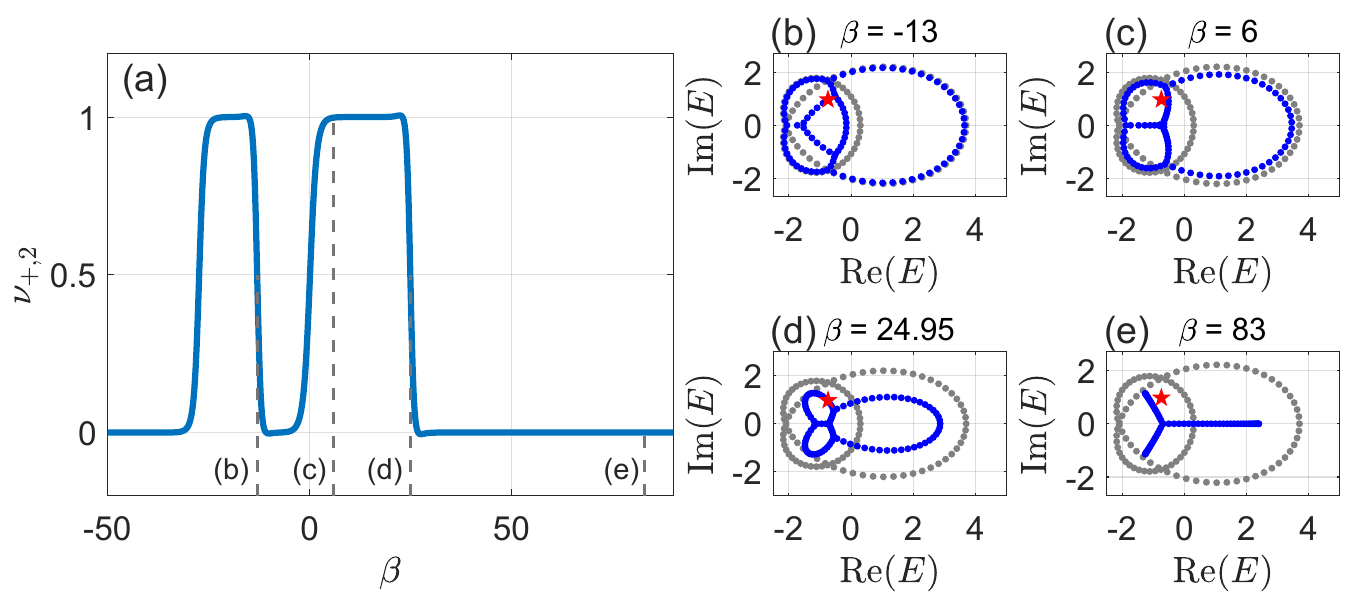}  
		\caption{Comparison between decompositions of quantized response and spectral winding for our model in the main text with spectral winding number $|W(E_r)|\leqslant2$.
		(a) Decomposition of quantized response quantity $\nu_{+,2}$, which forms two plateaus at $\nu_{+,2}=1$ for different regions of $\beta$. The two impurities are given by $\mu_2=e^{12}\mu_1=e^{12}e^{\beta}$, and the reference energy is $E_r=-0.74+0.96i$.
		(b) to (e) Blue dots show the spectra of the PBC system with impurities at different values of $\beta$ marked by the dash lines in (a). Gray dots show the PBC eigenenergies without impurities. Red stars mark the reference energy $E_r$.
		Other parameters are $t_1=1$, $t_{-1}=0.7$, $t_2=2$, $t_j=0$ for other $j$, and $N=100$.
}
		\label{sub:fig2}  

\end{figure}

In Fig. \ref{sub:fig2}, we compare these decomposed topologies under the same parameters.
It is seen in Fig. \ref{sub:fig2}(a) that the first plateau at $\nu_{+,2}=1$ ends at $\beta\approx -13$; whereas in the complex spectrum, the region with $W(E_r)=2$ has not yet vanished at this parameter, as shown in Fig. \ref{sub:fig2}(b).
Complete elimination of the large winding region occurs at $\beta\approx 6$, where the eigeneneriges forming the boundary of $W(E_r)=1$ region also deviate from their PBC counterparts, as shown in Fig. \ref{sub:fig2}(c). Consistently, this $\beta$ value falls in the second plateau of the quantized response, indicating that the second impurity (the weaker one) is also strong enough to significantly affect the system.
Similarly, as shown in Figs. \ref{sub:fig2}(d) and \ref{sub:fig2}(e) the second plateau ends at $\beta\approx 24.95$; whereas the regions with $W(E_r)=1$ has not vanished until $\beta\approx 83$.

% \vspace{-30pt}

\section{Topological decomposition with winding number $W(E_r)=3$}
In the main text, we ‌reveal the topological decomposition with two impurities for a specific example with $W(E_r)=2$. To further demonstrate the universality of this decomposition, we present the case in a similar model
with $W(E_r)=3$ and $N_{\mu}=3$ impurities in this supplemental section. Due to diverse impurity configurations, the decomposition displays richer behavior.

\begin{figure}[htbp]
		\includegraphics[width=0.9\linewidth]{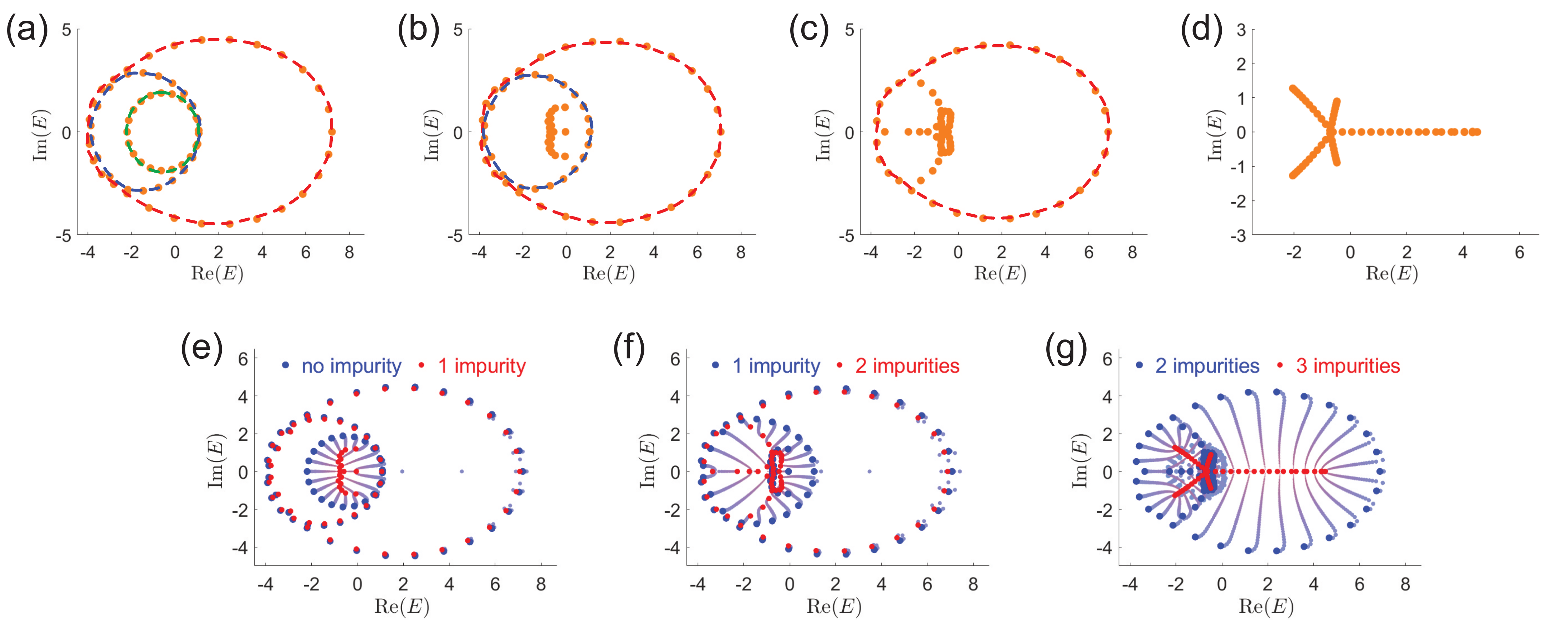}  
		\caption{Spectral winding and evolution with increasing impurity strengths.
		(a) Spectral winding for our model with $W(E_r)$ up to $3$ under PBCs in the absence of impurities, $\mu_1=\mu_2=\mu_3=0$. Orange dots are the spectrum on the complex energy plane. Red, blue, and gree dash lines correspond to the boundaries of regimes with $W(E_r)=1$, $W(E_r)=2$, and $W(E_r)=3$, respectively.
		(b) The same as in (a) but with $\mu_1=e^{25}$. Compared with (a), the green dash line shrinks into some curves and the region with $W(E_r)=3$ vanishes.
		(c) The same as in (a), but with $\mu_1=\mu_2=10^{26}$. Compared with (a), both the green and blue dash lines shrink into curves and the spectral winding number is now reduced to $W(E_r)=1$.
		(d) The same as in (a), but with $\mu_1=\mu_2=\mu_3=10^{35}$. The spectrum no longer encloses any area and $W(E_r)=0$ for arbitrary $E_r$, as under OBCs.
		(e) Spectral evolution (lilac color) between (a) and (b), with $\mu_2=\mu_3=0$ and $\mu_1$ increasing from $0$ (blue) to $e^{25}$ (red). Further increasing $\mu_1$ does not significantly affect the spectrum in the absence of additional impurities.
		(f) Spectral evolution between (b) and (c), but with $\mu_1=10^{26}$ and $\mu_2$ increasing from $0$ to $e^{26}$. 
		(g) Spectral evolution between (c) and (d), but with $\mu_1=\mu_2=10^{35}$ and $\mu_2$ increasing from $0$ to $e^{35}$. 
		Other parameters are $t_{1}=1$, $t_{-1}=0.7$, $t_{2}=2$,$t_{-2}=0.5$, $t_{3}=3$, $t_{-3}=0$,and $N=70$.
		Impurity states with exponentially large eigenenergies are omitted in all panels.
		}  
		\label{sub:fig3}  
\end{figure}

\subsection{Decomposition of non-Hermitian spectral topology}
In Fig. \ref{sub:fig3}, we display the decomposition of spectral winding topology with $W(E_r)=3$ and three impurity potentials $\mu_1$, $\mu_2$ and $\mu_3$, for the model of Eq. 1 of the main text.
Fig. \ref{sub:fig3}(a) demonstrates the PBC spectrum without any impurity, where segments of the spectrum marked by red, blue, and green dash lines correspond to the boundaries of areas with $W(E_r)=1$, $2$, and $3$, respectively.
By gradually increasing a single impurity, only the green one significantly shrinks and evolves into an $\varepsilon$-shaped curve that encloses no area [see Fig. \ref{sub:fig3}(e)], resulting in the spectrum in Fig. \ref{sub:fig3}(b) with $W(E_r)$ no larger than $2$.
Meanwhile, the red and blue segments in (a) remains roughly unchanged in this single-impurity case.
%As can be seen in Figs. \ref{sub:fig3}(a) and \ref{sub:fig3}(e), with increasing a single impurity $\mu_1$ on site $x=N$, the the region with $W(E_r)=3$ of PBC spectrum gradually evolves into its interior and reach the
%'$\epsilon$'-shaped spectrum in \ref{sub:fig3}(b). When $\mu_1$ exceeds a critical value, the region with $W(E_r)=3$ vanishes. Although the spectrum in \ref{sub:fig3}(b) appears to contain an internal area enclosed by '$\epsilon$'-shaped and loop-shaped spectrum and with a winding number $W(E_r)=3$, the winding direction for the '$\epsilon$'-shaped part of spectrum is not unidirectional. Specifically, the winding begins from lower end of '$\epsilon$' upward to the upper end, then retraces its path downward before reconnecting with the loop-shaped part of spectrum, returning to the starting point. Consequently, the '$\epsilon$'-shaped and the loop-shaped spectrum do not enclose any area, and the winding number of this spectrum is 2.
Next, by adding and increasing the second impurity $\mu_2$ at site $x=N-1$, 
the boundary of the region with $W(E_r)=2$ further shrinks into it interior [see Fig. \ref{sub:fig3}(f)] and eliminates this region, resulting in the spectrum in Fig. \ref{sub:fig3}(c) with $W(E_r)$ no larger than $1$. In comparison, only the red segment in (a) [boundary of $W(E_r)=1$] is not significantly affected in the presents of two strong impurities.
Finally, by adding and increasing the thrid impurity at site $x=N-2$, the full spectrum shrinks into the OBC spectrum without nontrivial spectral winding,  as shown in Figs. \ref{sub:fig3}(g) and \ref{sub:fig3}(d).

%By introducing and increasing the second impurity $\mu_2$ at site $x=N-1$, the region with $W(E_r)=2$ enclosed by loop-shaped spectrum start to shrinks, which can be seen in Figs. \ref{sub:fig3}(f),and reaches the '3'-shape spectrum in Figs. \ref{sub:fig3}(c) when $\mu_2$ exceeds a critical value. Analogous to the '$\epsilon$'-shaped scenario discussed above, the '3'-shape spectrum do not enclose any area and the winding number of this spectrum is 1.

%Finally, by introducing and increasing the third impurity $\mu_3$ on site $x=N-2$, the adjacent-loop spectrum shrinks into the OBC spectrum without nontrivial spectral winding, as shown in Figs. \ref{sub:fig3}(g) and \ref{sub:fig3}(d).

With the above results, we can see that the spectral winding topology with $W(E_r)=3$ is decomposed by sequentially adding the three impurities: a single impurity ($\mu_1$) reduces it to $W(E_r)=2$, an additional one  ($\mu_1$ and $\mu_2$) further reduces it to $W(E_r)=1$, and the presence of all the three impurities  ($\mu_1$, $\mu_2$, and $\mu_3$) eliminiate any nontrivial spectral winding.

%and reduced to $W(E_r)=2$ by a single on-site impurity ($\mu_1$), and further reduced to $W(E_r)=1$ by two on-site impurities ($\mu_1$and$\mu_2$). Then finally trivialized by three on-site impurities ($\mu_1$and$\mu_2$, $\mu_3$).
\subsection{Decomposition of non-Hermitian quantized response}
Next we demonstrate the decomposition of quantized steady-state response for $W(E_r)=3$.
%Beside the decomposition of non-Hermitian spectrum, the decomposition of quantized steady-state response also emerges. 
We consider the response quantity $\nu_{+,3}$ with fixed relative ratio $\eta_1$, $\eta_2$  between the amplitudes of the three impurities,
$$\mu_1=e^{\beta},~\mu_2=\eta_1\mu_1=\eta_1 e^\beta,~\mu_3={\eta_2}\mu_2={\eta_2}{\eta_1}\mu_1={\eta_2}{\eta_1}e^{\beta}.$$ 
%To simplify the impurity configuration, we first consider the case with ${\eta_2}={\eta_1}={\eta}$.
Note that the extra parameters $\eta_1$ and $\eta_2$ lead to richer impurity configurations and corresponding decomposition processes,
and we choose two explicit examples as demonstrations in the rest of this subsection.

\subsubsection{${\eta_2}={\eta_1}={\eta}$}
\begin{figure}[htbp]
		\includegraphics[width=0.9\linewidth]{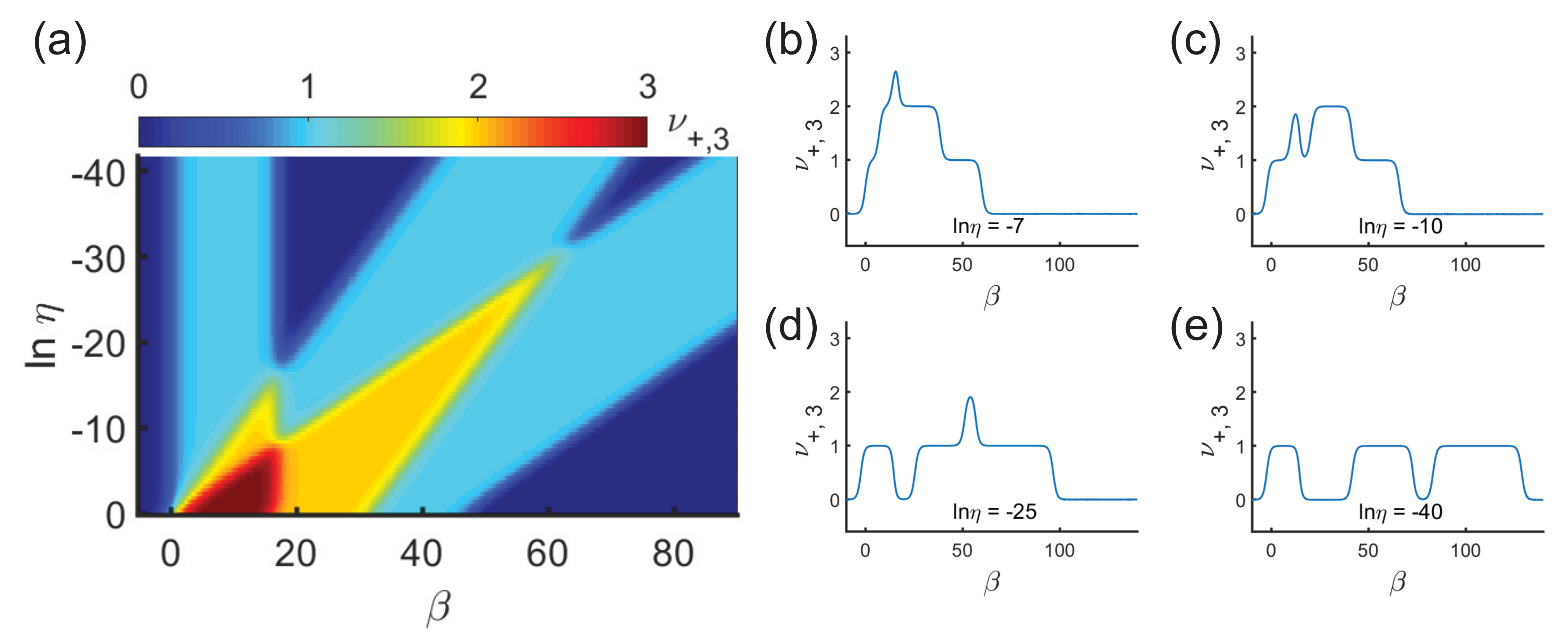}  
		\caption{Quantized response quantity $\nu_{+,3}$ for impurities $\mu_1=e^\beta,\mu_2=\eta e^\beta,\mu_3={\eta}^2 e^\beta,$ and reference energy $E_r= -1.5 + 0.5i$.
		(a) $\nu_{+,3}$ varying with $\beta$ for different ratio $\eta$ between three impurities.
		(b) to (e) Numerical (blue solid lines) results of the quantized response at $\ln c=-7,-10, -25, -40$, respectively. 
		Other parameters are $t_{1}=1$, $t_{-1}=0.7$, $t_{2}=2$,$t_{-2}=0.5$, $t_{3}=3$, $t_{-3}=0$,and $N=132$.
		}  
		\label{sub:fig4}  
\end{figure}
In this case, the strengths of the three impuries become
$$\mu_1=e^{\beta},~\mu_2=\eta e^\beta,~\mu_3={\eta^2}e^{\beta}.$$
Thus, when $\eta\ll1$, the system is expected to first enter the single-impurity scenario in Figs. \ref{sub:fig3}(b) and \ref{sub:fig3}(e) when
%a relatively small 
$1/\eta\gg e^\beta\gg 1$, 
then the double-impurity scenario in Figs. \ref{sub:fig3}(c) and \ref{sub:fig3}(f) when $1/\eta^2\gg\beta \gg1/\eta$, and the ‌triple-impurity scenario in Figs. \ref{sub:fig3}(d) and \ref{sub:fig3}(g) when $\beta\gg 1/\eta^2$.

The decomposition of quantized response quantity $\nu_{+,3}$ (see Eq. 3 in the main text for its definition) in this scenario is displayed in Fig. \ref{sub:fig4}.
When $\eta=1$, the three impurities are identical, resulting in a plateau at $\nu_{+,3}=3$ for small positive $\beta$, which stepwisely drops to $0$ when increasing $\beta$, as shown in Fig. \ref{sub:fig4}(a).
With $\eta$ decreased exponentially, the plateaus with $\nu_{+,3}=3$ (red) and $\nu_{+,3}=2$ (yellow) are gradually decomposed into those with $\nu_{+,3}=2$. Figs. \ref{sub:fig4}(b) to \ref{sub:fig4}(e) provide clear views of the decomposed plateaus at several specific values of $\eta$.
In particular, the quantized response for $W(E_r)=3$ is fully decomposed into three plateaus with $\nu_{+,3}=1$ for the chosen parameters with a system size $N=132$.

%Due to the complexity introduced by three impurities, plateau of $\nu_{+,3}=3$ exhibits more intricate decomposition as can be seen in Fig. \ref{sub:fig4}(a). With increasing the ratio $\eta$, the plateau of $\nu_{+,3}=3$ at $\eta=1$ is gradually decomposed into into two components: ‌neighboring $\nu_{+,3}=1$ and $\nu_{+,3}=2$ plateau pair at smaller $\beta$ and a plateau of $\nu_{+,3}=2$ at relatively larger $\beta$. With further increasing the ratio $\eta$, the plateau pair of $\nu_{+,3}=1$ and $\nu_{+,3}=2$ is reduced into one plateau of $\nu_{+,3}=1$, and the plateau of $\nu_{+,3}=2$ is gradually decomposed into into two plateaus of $\nu_{+,3}=1$.
%The complete decomposition occurs at a larger magnitude of $|\ln \eta |$, as indicated by the three fully separated plateaus of $\nu_{+,2}=1$. 
%In particular, the first plateau of smaller $\beta$ corresponds to the spectral evolution in Fig. \ref{sub:fig3}(e), the second one of medium $\beta$ corresponds to the spectral evolution in Fig. \ref{sub:fig3}(f), and the third one of larger $\beta$ corresponds to the spectral evolution in Fig. \ref{sub:fig3}(g). Plateaus drop to zero when eigenenergies pass through the reference energy $E_r$ during the evolution.
\subsubsection{${\eta_2}=\eta,~{\eta_1}=1$}
\begin{figure}[htbp]
		\includegraphics[width=0.9\linewidth]{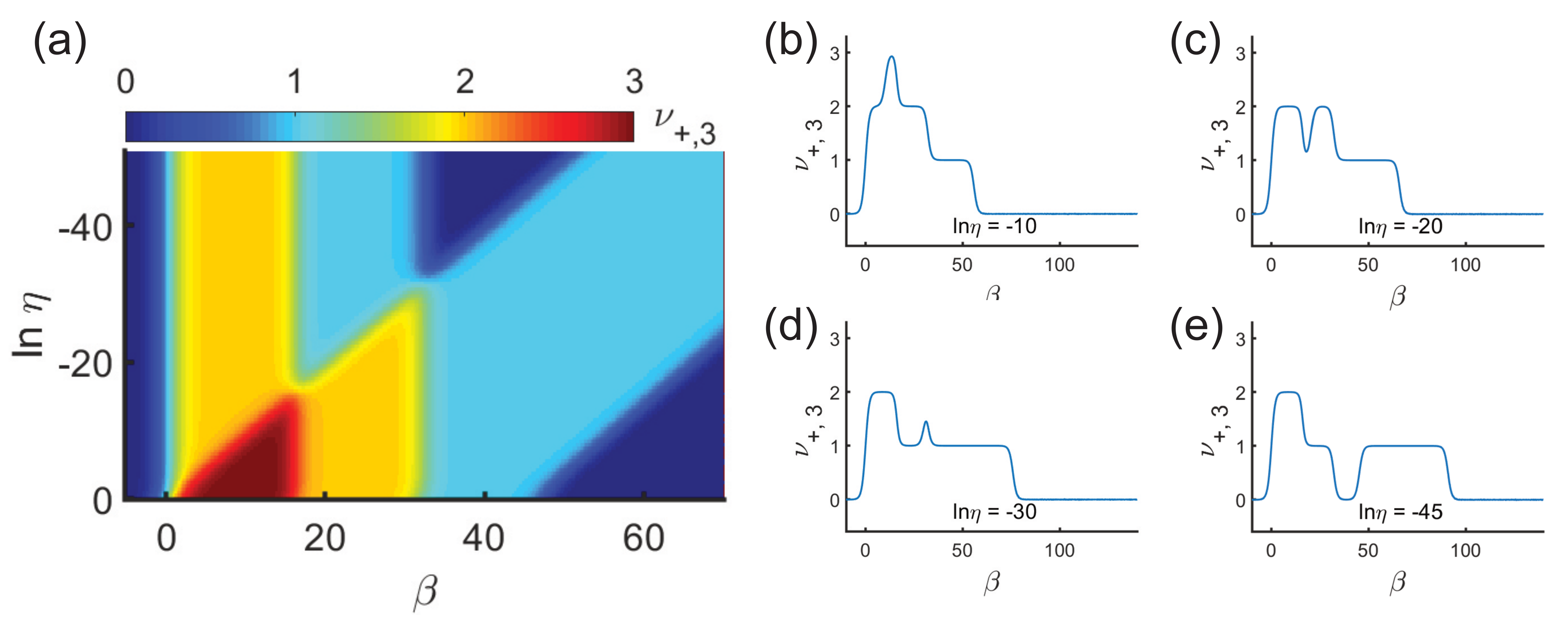}  
		\caption{Quantized response quantity $\nu_{+,3}$ for impurities 
		$\mu_1=\mu_2=e^\beta,\mu_3={\eta}e^{\beta},$ and reference energy $E_r= -1.5 + 0.5i$.
		(a) $\nu_{+,3}$ varying with $\beta$ for different ratio $\eta$ between three impurities. The plateau of $\nu_{+,3}=3$ is decomposed to two of $\nu_{+,3}=2$ when $|\ln \eta|$ increases.When $|\ln \eta|$ increases further, the plateau of $\nu_{+,3}=2$ at larger $\beta$ is decomposed to two of $\nu_{+,3}=1$.
		(b) to (e) Numerical (blue solid lines) results of the quantized response at $\ln \eta=-10,-20, -30, -45$, respectively. 
		Other parameters are $t_{1}=1$, $t_{-1}=0.7$, $t_{2}=2$,$t_{-2}=0.5$, $t_{3}=3$, $t_{-3}=0$,and $N=132$.
		}  
		\label{sub:fig5}  
\end{figure}
Next we consider the case with two identical impurities,
$$\mu_1=\mu_2=e^\beta,~\mu_3=\eta e^\beta.$$
%For different impurity configurations, the quantum steady-state response exhibits a rich and diverse decomposition processes. Here we consider another impurity configuration with ${\eta_1}=1$.
As can be seen from Fig. \ref{sub:fig5}(a), the plateau of $W(E_r)=3$ at $\eta=1$ is decomposed into two with $W(E_r)=2$ and $W(E_r)=1$ when $\ln\eta\lesssim-30$. Note that each plateau mentioned here stepwisely drops with decreasing $\beta$, reflecting the reduction of spectral winding topology when impurities get stronger.
In particular, the plateau of $W(E_r)=2$ (and its stepwisely dropping to $0$) remains unchanged with $\eta$ further decreasing beyond approximately $e^{-30}$, reflecting the influence of the two $\eta$-independent impurities $\mu_1$ and $\mu_2$; and the other plateau of $W(E_r)=1$ moves along $\beta$-axis when decreasing $\eta$, reflecting the influence of the third impurity $\mu_3=\eta e^\beta$.
%the plateau of $\nu_{+,3}=3$ at $\eta=1$ is gradually decomposed into two plateau of $\nu_{+,3}=2$ as can be seen in Fig. \ref{sub:fig5}(b) and \ref{sub:fig5}(c). With further increasing the ratio $\eta$, the plateau of $\nu_{+,3}=2$ at larger $\beta$ is decomposed to two of $\nu_{+,3}=1$ as can be seen in Fig. \ref{sub:fig5}(d) and \ref{sub:fig5}(e).
%For our chosen parameters, the plateau of $\nu_{+,3}=2$ at smaller $\beta$ remains and do not decompose.
Figs. \ref{sub:fig5}(b) to \ref{sub:fig5}(e) provide clear views of the decomposed plateaus at several specific values of $\eta$.

\subsection{Decomposition of Hermitian band topology}
\begin{figure}[htbp]
		\includegraphics[width=0.9\linewidth]{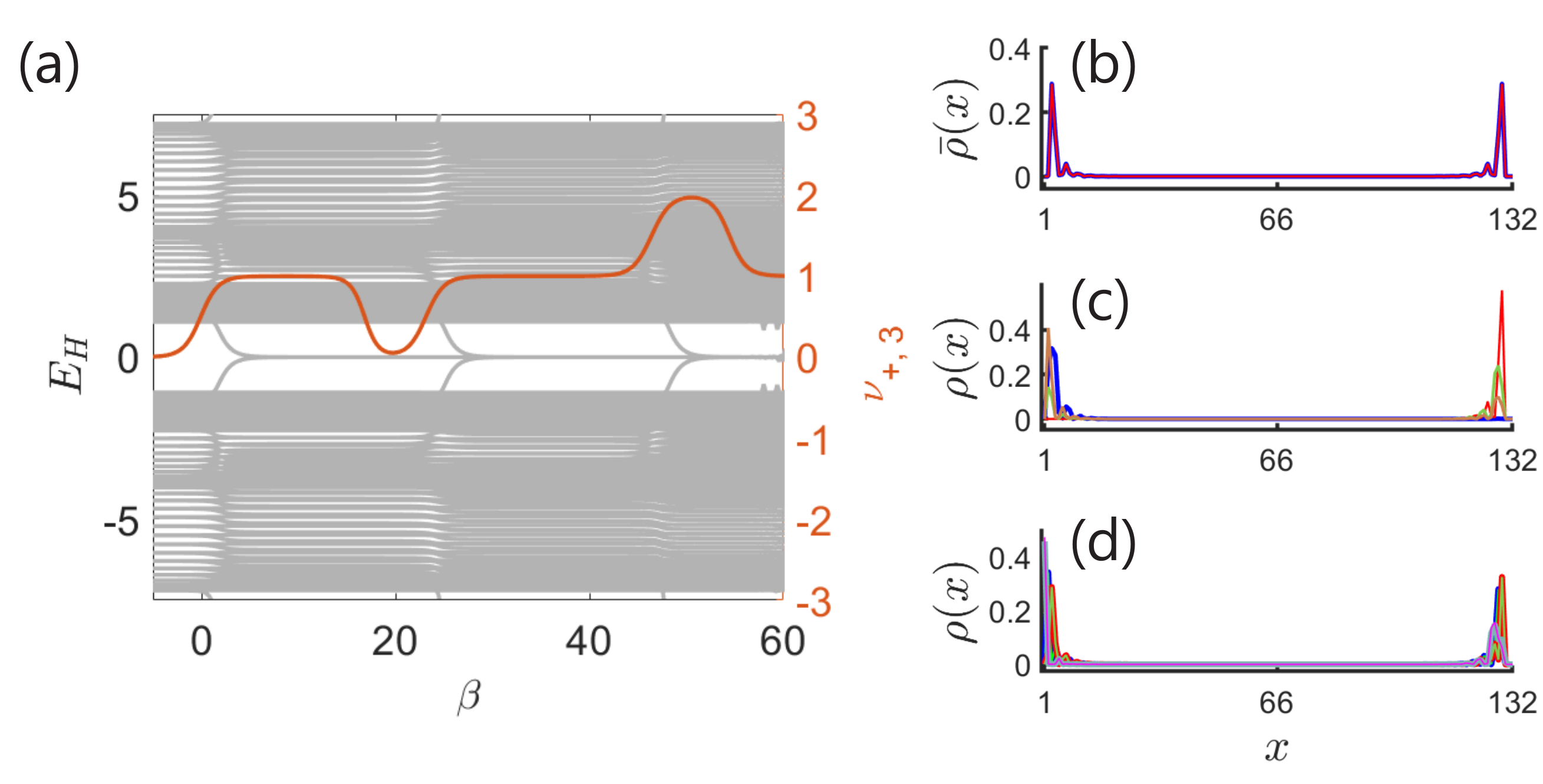}  
		\caption{Topological properties of the the doubled Hermitian Hamiltonian $H_{\rm H}$
		(a) OBC spectrum of $H_{\rm H}$ with $E_r=-1.5+0.5i$ and $W(E_r)=3$, in comparison of the quantized response quantity $\nu_{+,3}$ for the non-Hermitian Hamiltonian $H$.
		A pair of zero-energy edge states appears when $\nu_{+,3}$ increases by $1$.
		(b) to (d) Distribution of  topological edge states at $\beta\approx 10, 30$ and $50$. ${\rho}(x)$ represents the density of eigenstates summed over the two pseudospin components of $H_{\rm H}$.
		Other parameters are $t_1=1$, $t_{-1}=0.7$, $t_2=2$, $t_{-2}=0.5$, $t_3=3$, $t_{-3}=0$, $\mu_3={\eta}\mu_2={\eta}^2\mu_1={\eta}^2e^{\beta}$ where $\eta=e^{-23}$, and $N=132$.
		}
		\label{sub:fig6}  
\end{figure}

In this subsection, we demonstrate numerical results of the doubled Hermitian 
$
{H}_{\rm H}=
\begin{pmatrix}
0&H-E_r\\
H^\dagger-E_r^*&0
\end{pmatrix}
$,
with $H$ being the non-Hermitian Hamiltonian with $W(E_r)=3$ discussed in the last subsection, in comparison to the quantized response quantity $\nu_{+,3}$ of $H$.
%for $W(E_r)=3$ with strongly detuned impurities ($\eta=e^{-23}$),  in comparison to the  quantized response quantity $\nu_{+,3}$ of the non-Hermitian Hamiltonian $H$.
In particular, Fig. \ref{sub:fig6}(a) demonstrates the OBC spectra of  ${H}_{\rm H}$ with three distinguished impurities, $\mu_1=e^\beta$, $\mu_2=\eta e^\beta$, and $\mu_3=\eta^2 e^\beta$.
With increasing $\beta$, three pairs of zero-energy edge states emerge at different values of $\beta$, in consistence with the decomposed quantized plateaus of $\nu_{+,3}$.
The distribution of these zero-energy edge states are shown in Figs. \ref{sub:fig6}(b) to (d).
Note that here the results are obtained numerically for a finite-size system, where $\nu_{+,3}$ inevitably drops at some size-dependent values of $\beta$. As discussed in the main text, this cannot be reflected by the zero-energy edge states that represent a size-insensitive feature of Hermitian topological phases; and in turn, the number of these states cannot be directly reflected by $\nu_{+,3}$.
Nevertheless, the correspondence between Hermitian topological edge states and non-Hermitian quantized response can be established 
as follow: a pair of zero-energy edge states emerges for $H_{\rm H}$ whenever the quantized quantity $\nu_{+,3}$ increases by $1$.

%By increasing $\beta$, the impurities $\mu_1$, $\mu_2$ and $\mu_3$ induce impurity states when $\beta$ approximately exceed $0$, $20$ and $45$, respectively, whose eigenenergies exponentially growing with $\beta$. At each of these critical values of $\beta$, the quantized response quantity $\nu^{\rm ana}_{+,3}$ increases by $1$, and a pair of zero-energy edge states emerge from the bulk bands of $H_{\rm H}$. Distributions of these impurity and zero-energy edge states are illustrated in Fig. \ref{fig4}(b) to \ref{fig4}(d). Each impurity induces one pair of topological zero-energy states to $H_{\rm H}$, representing a decomposition of the Hermitian topology with $W(E_r)=3$.

\begin{figure}[H]
		\includegraphics[width=0.92\linewidth]{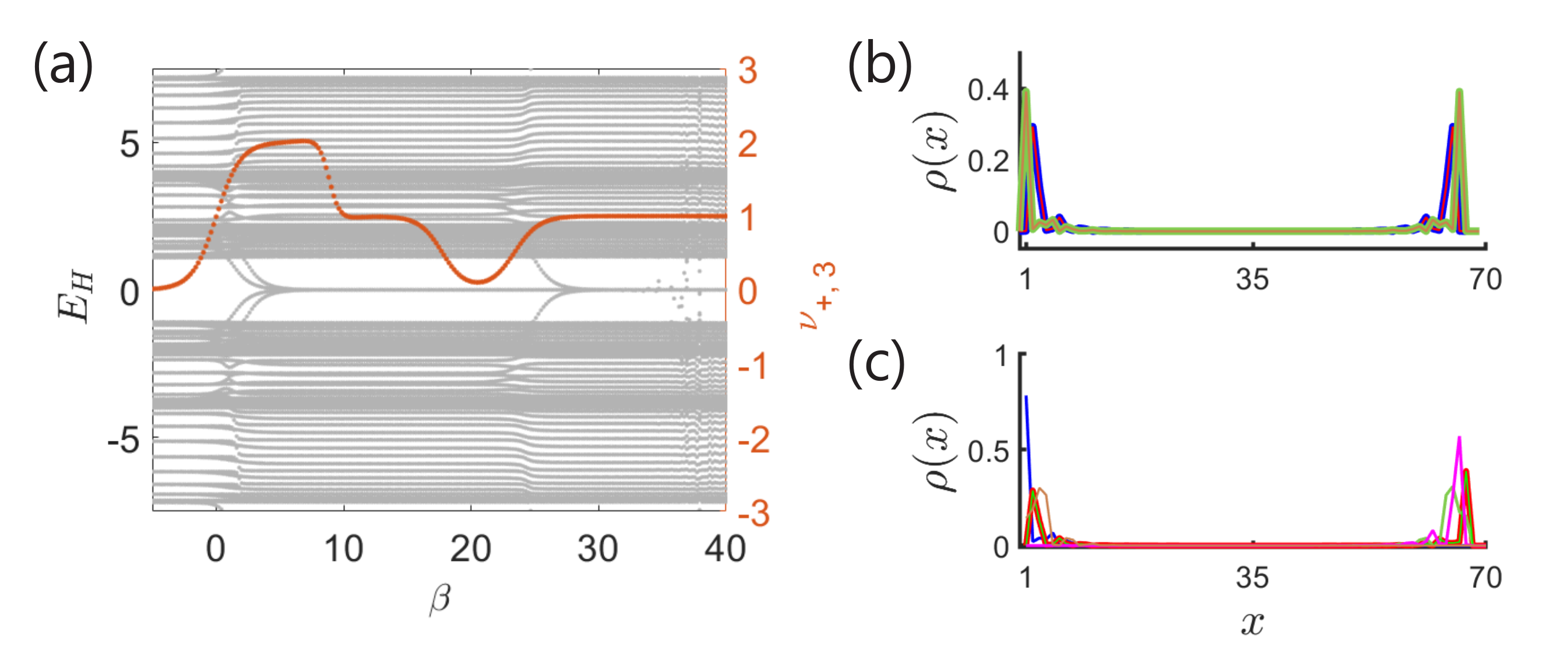}  
		\caption{Topological properties of the the doubled Hermitian Hamiltonian $H_{\rm H}$
		(a) OBC spectrum of $H_{\rm H}$ with $E_r=-1.5+0.5i$ and $W(E_r)=3$, in comparison of the quantized response quantity $\nu_{+,3}$ for the non-Hermitian Hamiltonian $H$.
		Two pairs of zero-energy edge states appear when $\nu_{+,3}$ increases by $2$ at $\beta\approx 0$, and another pair of zero-energy edge states appears when $\nu_{+,3}$ increases by $1$ at $\beta\approx 25$.
		(b) to (c) Distribution of impurity and topological edge states at $\beta\approx 15 $ and $30$. ${\rho}(x)$ represents the density of eigenstates summed over the two pseudospin components of $H_{\rm H}$.
		Other parameters are $t_1=1$, $t_{-1}=0.7$, $t_2=2$, $t_{-2}=0.5$, $t_3=3$, $t_{-3}=0$, $\mu_3={\eta}\mu_2={\eta}\mu_1={\eta}e^{\beta}$ where $\eta=10^{-23}$, and $N=70$.
		}
		\label{sub:fig7}  
\end{figure}

In Fig. \ref{sub:fig7},
we display the results of $H_{\rm H}$ for the other case of impurities discussed in the last subsection, with $\mu_1=\mu_2=e^\beta$ and $\mu_3=\eta e^\beta$.
%another example with $W(E_r)=3$ as a comparison is displayed. 
Similarly to the previous case, two pairs of zero-energy edge states appear when $\nu_{+,3}$ increases by $2$ at $\beta\approx 0 $, and another pair of zero-energy edge states appear when $\nu_{+,3}$ increases by $1$ at $\beta\approx 25$.

\clearpage
\bibliography{ref_topoDecomposition.bib}

%merlin.mbs apsrev4-1.bst 2010-07-25 4.21a (PWD, AO, DPC) hacked
%Control: key (0)
%Control: author (0) dotless jnrlst
%Control: editor formatted (1) identically to author
%Control: production of article title (0) allowed
%Control: page (1) range
%Control: year (0) verbatim
%Control: production of eprint (0) enabled
\begin{thebibliography}{41}%
\makeatletter
\providecommand \@ifxundefined [1]{%
 \@ifx{#1\undefined}
}%
\providecommand \@ifnum [1]{%
 \ifnum #1\expandafter \@firstoftwo
 \else \expandafter \@secondoftwo
 \fi
}%
\providecommand \@ifx [1]{%
 \ifx #1\expandafter \@firstoftwo
 \else \expandafter \@secondoftwo
 \fi
}%
\providecommand \natexlab [1]{#1}%
\providecommand \enquote  [1]{``#1''}%
\providecommand \bibnamefont  [1]{#1}%
\providecommand \bibfnamefont [1]{#1}%
\providecommand \citenamefont [1]{#1}%
\providecommand \href@noop [0]{\@secondoftwo}%
\providecommand \href [0]{\begingroup \@sanitize@url \@href}%
\providecommand \@href[1]{\@@startlink{#1}\@@href}%
\providecommand \@@href[1]{\endgroup#1\@@endlink}%
\providecommand \@sanitize@url [0]{\catcode `\\12\catcode `\$12\catcode
  `\&12\catcode `\#12\catcode `\^12\catcode `\_12\catcode `\%12\relax}%
\providecommand \@@startlink[1]{}%
\providecommand \@@endlink[0]{}%
\providecommand \url  [0]{\begingroup\@sanitize@url \@url }%
\providecommand \@url [1]{\endgroup\@href {#1}{\urlprefix }}%
\providecommand \urlprefix  [0]{URL }%
\providecommand \Eprint [0]{\href }%
\providecommand \doibase [0]{http://dx.doi.org/}%
\providecommand \selectlanguage [0]{\@gobble}%
\providecommand \bibinfo  [0]{\@secondoftwo}%
\providecommand \bibfield  [0]{\@secondoftwo}%
\providecommand \translation [1]{[#1]}%
\providecommand \BibitemOpen [0]{}%
\providecommand \bibitemStop [0]{}%
\providecommand \bibitemNoStop [0]{.\EOS\space}%
\providecommand \EOS [0]{\spacefactor3000\relax}%
\providecommand \BibitemShut  [1]{\csname bibitem#1\endcsname}%
\let\auto@bib@innerbib\@empty
%</preamble>
\bibitem [{\citenamefont {Hasan}\ and\ \citenamefont
  {Kane}(2010)}]{hasan2010colloquium}%
  \BibitemOpen
  \bibfield  {author} {\bibinfo {author} {\bibfnamefont {M~Zahid}\ \bibnamefont
  {Hasan}}\ and\ \bibinfo {author} {\bibfnamefont {Charles~L}\ \bibnamefont
  {Kane}},\ }\bibfield  {title} {\enquote {\bibinfo {title} {Colloquium:
  topological insulators},}\ }\href@noop {} {\bibfield  {journal} {\bibinfo
  {journal} {Reviews of modern physics}\ }\textbf {\bibinfo {volume} {82}},\
  \bibinfo {pages} {3045--3067} (\bibinfo {year} {2010})}\BibitemShut {NoStop}%
\bibitem [{\citenamefont {Qi}\ and\ \citenamefont
  {Zhang}(2011)}]{qi2011topological}%
  \BibitemOpen
  \bibfield  {author} {\bibinfo {author} {\bibfnamefont {Xiao-Liang}\
  \bibnamefont {Qi}}\ and\ \bibinfo {author} {\bibfnamefont {Shou-Cheng}\
  \bibnamefont {Zhang}},\ }\bibfield  {title} {\enquote {\bibinfo {title}
  {Topological insulators and superconductors},}\ }\href@noop {} {\bibfield
  {journal} {\bibinfo  {journal} {Reviews of modern physics}\ }\textbf
  {\bibinfo {volume} {83}},\ \bibinfo {pages} {1057--1110} (\bibinfo {year}
  {2011})}\BibitemShut {NoStop}%
\bibitem [{\citenamefont {Bansil}\ \emph {et~al.}(2016)\citenamefont {Bansil},
  \citenamefont {Lin},\ and\ \citenamefont {Das}}]{RevModPhys.88.021004}%
  \BibitemOpen
  \bibfield  {author} {\bibinfo {author} {\bibfnamefont {A.}~\bibnamefont
  {Bansil}}, \bibinfo {author} {\bibfnamefont {Hsin}\ \bibnamefont {Lin}}, \
  and\ \bibinfo {author} {\bibfnamefont {Tanmoy}\ \bibnamefont {Das}},\
  }\bibfield  {title} {\enquote {\bibinfo {title} {Colloquium: Topological band
  theory},}\ }\href {\doibase 10.1103/RevModPhys.88.021004} {\bibfield
  {journal} {\bibinfo  {journal} {Rev. Mod. Phys.}\ }\textbf {\bibinfo {volume}
  {88}},\ \bibinfo {pages} {021004} (\bibinfo {year} {2016})}\BibitemShut
  {NoStop}%
\bibitem [{\citenamefont {Chiu}\ \emph {et~al.}(2016)\citenamefont {Chiu},
  \citenamefont {Teo}, \citenamefont {Schnyder},\ and\ \citenamefont
  {Ryu}}]{RevModPhys.88.035005}%
  \BibitemOpen
  \bibfield  {author} {\bibinfo {author} {\bibfnamefont {Ching-Kai}\
  \bibnamefont {Chiu}}, \bibinfo {author} {\bibfnamefont {Jeffrey C.~Y.}\
  \bibnamefont {Teo}}, \bibinfo {author} {\bibfnamefont {Andreas~P.}\
  \bibnamefont {Schnyder}}, \ and\ \bibinfo {author} {\bibfnamefont {Shinsei}\
  \bibnamefont {Ryu}},\ }\bibfield  {title} {\enquote {\bibinfo {title}
  {Classification of topological quantum matter with symmetries},}\ }\href
  {\doibase 10.1103/RevModPhys.88.035005} {\bibfield  {journal} {\bibinfo
  {journal} {Rev. Mod. Phys.}\ }\textbf {\bibinfo {volume} {88}},\ \bibinfo
  {pages} {035005} (\bibinfo {year} {2016})}\BibitemShut {NoStop}%
\bibitem [{\citenamefont {Xie}\ \emph {et~al.}(2021)\citenamefont {Xie},
  \citenamefont {Wang}, \citenamefont {Zhang}, \citenamefont {Zhan},
  \citenamefont {Jiang}, \citenamefont {Lu},\ and\ \citenamefont
  {Chen}}]{xie2021higher}%
  \BibitemOpen
  \bibfield  {author} {\bibinfo {author} {\bibfnamefont {Biye}\ \bibnamefont
  {Xie}}, \bibinfo {author} {\bibfnamefont {Hai-Xiao}\ \bibnamefont {Wang}},
  \bibinfo {author} {\bibfnamefont {Xiujuan}\ \bibnamefont {Zhang}}, \bibinfo
  {author} {\bibfnamefont {Peng}\ \bibnamefont {Zhan}}, \bibinfo {author}
  {\bibfnamefont {Jian-Hua}\ \bibnamefont {Jiang}}, \bibinfo {author}
  {\bibfnamefont {Minghui}\ \bibnamefont {Lu}}, \ and\ \bibinfo {author}
  {\bibfnamefont {Yanfeng}\ \bibnamefont {Chen}},\ }\bibfield  {title}
  {\enquote {\bibinfo {title} {Higher-order band topology},}\ }\href@noop {}
  {\bibfield  {journal} {\bibinfo  {journal} {Nature Reviews Physics}\ }\textbf
  {\bibinfo {volume} {3}},\ \bibinfo {pages} {520--532} (\bibinfo {year}
  {2021})}\BibitemShut {NoStop}%
\bibitem [{\citenamefont {Gong}\ \emph {et~al.}(2018)\citenamefont {Gong},
  \citenamefont {Ashida}, \citenamefont {Kawabata}, \citenamefont {Takasan},
  \citenamefont {Higashikawa},\ and\ \citenamefont
  {Ueda}}]{gong2018topological}%
  \BibitemOpen
  \bibfield  {author} {\bibinfo {author} {\bibfnamefont {Zongping}\
  \bibnamefont {Gong}}, \bibinfo {author} {\bibfnamefont {Yuto}\ \bibnamefont
  {Ashida}}, \bibinfo {author} {\bibfnamefont {Kohei}\ \bibnamefont
  {Kawabata}}, \bibinfo {author} {\bibfnamefont {Kazuaki}\ \bibnamefont
  {Takasan}}, \bibinfo {author} {\bibfnamefont {Sho}\ \bibnamefont
  {Higashikawa}}, \ and\ \bibinfo {author} {\bibfnamefont {Masahito}\
  \bibnamefont {Ueda}},\ }\bibfield  {title} {\enquote {\bibinfo {title}
  {Topological phases of non-hermitian systems},}\ }\href@noop {} {\bibfield
  {journal} {\bibinfo  {journal} {Physical Review X}\ }\textbf {\bibinfo
  {volume} {8}},\ \bibinfo {pages} {031079} (\bibinfo {year}
  {2018})}\BibitemShut {NoStop}%
\bibitem [{\citenamefont {Okuma}\ \emph {et~al.}(2020)\citenamefont {Okuma},
  \citenamefont {Kawabata}, \citenamefont {Shiozaki},\ and\ \citenamefont
  {Sato}}]{okuma2020topological}%
  \BibitemOpen
  \bibfield  {author} {\bibinfo {author} {\bibfnamefont {Nobuyuki}\
  \bibnamefont {Okuma}}, \bibinfo {author} {\bibfnamefont {Kohei}\ \bibnamefont
  {Kawabata}}, \bibinfo {author} {\bibfnamefont {Ken}\ \bibnamefont
  {Shiozaki}}, \ and\ \bibinfo {author} {\bibfnamefont {Masatoshi}\
  \bibnamefont {Sato}},\ }\bibfield  {title} {\enquote {\bibinfo {title}
  {Topological origin of non-hermitian skin effects},}\ }\href@noop {}
  {\bibfield  {journal} {\bibinfo  {journal} {Physical review letters}\
  }\textbf {\bibinfo {volume} {124}},\ \bibinfo {pages} {086801} (\bibinfo
  {year} {2020})}\BibitemShut {NoStop}%
\bibitem [{\citenamefont {Zhang}\ \emph {et~al.}(2020)\citenamefont {Zhang},
  \citenamefont {Yang},\ and\ \citenamefont {Fang}}]{zhang2020correspondence}%
  \BibitemOpen
  \bibfield  {author} {\bibinfo {author} {\bibfnamefont {Kai}\ \bibnamefont
  {Zhang}}, \bibinfo {author} {\bibfnamefont {Zhesen}\ \bibnamefont {Yang}}, \
  and\ \bibinfo {author} {\bibfnamefont {Chen}\ \bibnamefont {Fang}},\
  }\bibfield  {title} {\enquote {\bibinfo {title} {Correspondence between
  winding numbers and skin modes in non-hermitian systems},}\ }\href@noop {}
  {\bibfield  {journal} {\bibinfo  {journal} {Physical Review Letters}\
  }\textbf {\bibinfo {volume} {125}},\ \bibinfo {pages} {126402} (\bibinfo
  {year} {2020})}\BibitemShut {NoStop}%
\bibitem [{\citenamefont {Wang}\ \emph {et~al.}(2021)\citenamefont {Wang},
  \citenamefont {Dutt}, \citenamefont {Yang}, \citenamefont {Wojcik},
  \citenamefont {Vu{\v{c}}kovi{\'c}},\ and\ \citenamefont
  {Fan}}]{wang2021generating}%
  \BibitemOpen
  \bibfield  {author} {\bibinfo {author} {\bibfnamefont {Kai}\ \bibnamefont
  {Wang}}, \bibinfo {author} {\bibfnamefont {Avik}\ \bibnamefont {Dutt}},
  \bibinfo {author} {\bibfnamefont {Ki~Youl}\ \bibnamefont {Yang}}, \bibinfo
  {author} {\bibfnamefont {Casey~C}\ \bibnamefont {Wojcik}}, \bibinfo {author}
  {\bibfnamefont {Jelena}\ \bibnamefont {Vu{\v{c}}kovi{\'c}}}, \ and\ \bibinfo
  {author} {\bibfnamefont {Shanhui}\ \bibnamefont {Fan}},\ }\bibfield  {title}
  {\enquote {\bibinfo {title} {Generating arbitrary topological windings of a
  non-hermitian band},}\ }\href@noop {} {\bibfield  {journal} {\bibinfo
  {journal} {Science}\ }\textbf {\bibinfo {volume} {371}},\ \bibinfo {pages}
  {1240--1245} (\bibinfo {year} {2021})}\BibitemShut {NoStop}%
\bibitem [{\citenamefont {Zhang}\ \emph {et~al.}(2021)\citenamefont {Zhang},
  \citenamefont {Yang}, \citenamefont {Ge}, \citenamefont {Guan}, \citenamefont
  {Chen}, \citenamefont {Yan}, \citenamefont {Chen}, \citenamefont {Xi},
  \citenamefont {Li}, \citenamefont {Jia} \emph {et~al.}}]{zhang2021acoustic}%
  \BibitemOpen
  \bibfield  {author} {\bibinfo {author} {\bibfnamefont {Li}~\bibnamefont
  {Zhang}}, \bibinfo {author} {\bibfnamefont {Yihao}\ \bibnamefont {Yang}},
  \bibinfo {author} {\bibfnamefont {Yong}\ \bibnamefont {Ge}}, \bibinfo
  {author} {\bibfnamefont {Yi-Jun}\ \bibnamefont {Guan}}, \bibinfo {author}
  {\bibfnamefont {Qiaolu}\ \bibnamefont {Chen}}, \bibinfo {author}
  {\bibfnamefont {Qinghui}\ \bibnamefont {Yan}}, \bibinfo {author}
  {\bibfnamefont {Fujia}\ \bibnamefont {Chen}}, \bibinfo {author}
  {\bibfnamefont {Rui}\ \bibnamefont {Xi}}, \bibinfo {author} {\bibfnamefont
  {Yuanzhen}\ \bibnamefont {Li}}, \bibinfo {author} {\bibfnamefont {Ding}\
  \bibnamefont {Jia}},  \emph {et~al.},\ }\bibfield  {title} {\enquote
  {\bibinfo {title} {Acoustic non-hermitian skin effect from twisted winding
  topology},}\ }\href@noop {} {\bibfield  {journal} {\bibinfo  {journal}
  {Nature communications}\ }\textbf {\bibinfo {volume} {12}},\ \bibinfo {pages}
  {6297} (\bibinfo {year} {2021})}\BibitemShut {NoStop}%
\bibitem [{\citenamefont {Su}\ \emph {et~al.}(2021)\citenamefont {Su},
  \citenamefont {Estrecho}, \citenamefont {Biega{\'n}ska}, \citenamefont
  {Huang}, \citenamefont {Wurdack}, \citenamefont {Pieczarka}, \citenamefont
  {Truscott}, \citenamefont {Liew}, \citenamefont {Ostrovskaya},\ and\
  \citenamefont {Xiong}}]{su2021direct}%
  \BibitemOpen
  \bibfield  {author} {\bibinfo {author} {\bibfnamefont {Rui}\ \bibnamefont
  {Su}}, \bibinfo {author} {\bibfnamefont {Eliezer}\ \bibnamefont {Estrecho}},
  \bibinfo {author} {\bibfnamefont {D\c{a}br{\'o}wka}\ \bibnamefont
  {Biega{\'n}ska}}, \bibinfo {author} {\bibfnamefont {Yuqing}\ \bibnamefont
  {Huang}}, \bibinfo {author} {\bibfnamefont {Matthias}\ \bibnamefont
  {Wurdack}}, \bibinfo {author} {\bibfnamefont {Maciej}\ \bibnamefont
  {Pieczarka}}, \bibinfo {author} {\bibfnamefont {Andrew~G}\ \bibnamefont
  {Truscott}}, \bibinfo {author} {\bibfnamefont {Timothy~CH}\ \bibnamefont
  {Liew}}, \bibinfo {author} {\bibfnamefont {Elena~A}\ \bibnamefont
  {Ostrovskaya}}, \ and\ \bibinfo {author} {\bibfnamefont {Qihua}\ \bibnamefont
  {Xiong}},\ }\bibfield  {title} {\enquote {\bibinfo {title} {Direct
  measurement of a non-hermitian topological invariant in a hybrid light-matter
  system},}\ }\href@noop {} {\bibfield  {journal} {\bibinfo  {journal} {Science
  advances}\ }\textbf {\bibinfo {volume} {7}},\ \bibinfo {pages} {eabj8905}
  (\bibinfo {year} {2021})}\BibitemShut {NoStop}%
\bibitem [{\citenamefont {Cao}\ \emph {et~al.}(2023)\citenamefont {Cao},
  \citenamefont {Li}, \citenamefont {Zhao}, \citenamefont {Guo}, \citenamefont
  {Qi}, \citenamefont {Chang}, \citenamefont {Zhou}, \citenamefont {Xu},\ and\
  \citenamefont {Duan}}]{cao2023probing}%
  \BibitemOpen
  \bibfield  {author} {\bibinfo {author} {\bibfnamefont {M-M}\ \bibnamefont
  {Cao}}, \bibinfo {author} {\bibfnamefont {Kai}\ \bibnamefont {Li}}, \bibinfo
  {author} {\bibfnamefont {W-D}\ \bibnamefont {Zhao}}, \bibinfo {author}
  {\bibfnamefont {W-X}\ \bibnamefont {Guo}}, \bibinfo {author} {\bibfnamefont
  {B-X}\ \bibnamefont {Qi}}, \bibinfo {author} {\bibfnamefont {X-Y}\
  \bibnamefont {Chang}}, \bibinfo {author} {\bibfnamefont {Z-C}\ \bibnamefont
  {Zhou}}, \bibinfo {author} {\bibfnamefont {Yong}\ \bibnamefont {Xu}}, \ and\
  \bibinfo {author} {\bibfnamefont {L-M}\ \bibnamefont {Duan}},\ }\bibfield
  {title} {\enquote {\bibinfo {title} {Probing complex-energy topology via
  non-hermitian absorption spectroscopy in a trapped ion simulator},}\
  }\href@noop {} {\bibfield  {journal} {\bibinfo  {journal} {Physical Review
  Letters}\ }\textbf {\bibinfo {volume} {130}},\ \bibinfo {pages} {163001}
  (\bibinfo {year} {2023})}\BibitemShut {NoStop}%
\bibitem [{\citenamefont {Wang}\ \emph {et~al.}(2024)\citenamefont {Wang},
  \citenamefont {Zhong},\ and\ \citenamefont {Fan}}]{wang2024non}%
  \BibitemOpen
  \bibfield  {author} {\bibinfo {author} {\bibfnamefont {Heming}\ \bibnamefont
  {Wang}}, \bibinfo {author} {\bibfnamefont {Janet}\ \bibnamefont {Zhong}}, \
  and\ \bibinfo {author} {\bibfnamefont {Shanhui}\ \bibnamefont {Fan}},\
  }\bibfield  {title} {\enquote {\bibinfo {title} {Non-hermitian photonic band
  winding and skin effects: a tutorial},}\ }\href@noop {} {\bibfield  {journal}
  {\bibinfo  {journal} {Advances in Optics and Photonics}\ }\textbf {\bibinfo
  {volume} {16}},\ \bibinfo {pages} {659--748} (\bibinfo {year}
  {2024})}\BibitemShut {NoStop}%
\bibitem [{\citenamefont {Yang}\ \emph {et~al.}(2025)\citenamefont {Yang},
  \citenamefont {Liao}, \citenamefont {Zhang}, \citenamefont {Li},
  \citenamefont {Hao}, \citenamefont {Zhou}, \citenamefont {Luo}, \citenamefont
  {Xu}, \citenamefont {Li},\ and\ \citenamefont {Guo}}]{yang2025observing}%
  \BibitemOpen
  \bibfield  {author} {\bibinfo {author} {\bibfnamefont {Mu}~\bibnamefont
  {Yang}}, \bibinfo {author} {\bibfnamefont {Yu-Wei}\ \bibnamefont {Liao}},
  \bibinfo {author} {\bibfnamefont {Hao-Qing}\ \bibnamefont {Zhang}}, \bibinfo
  {author} {\bibfnamefont {Yue}\ \bibnamefont {Li}}, \bibinfo {author}
  {\bibfnamefont {Zhi-He}\ \bibnamefont {Hao}}, \bibinfo {author}
  {\bibfnamefont {Zheng-Wei}\ \bibnamefont {Zhou}}, \bibinfo {author}
  {\bibfnamefont {Xi-Wang}\ \bibnamefont {Luo}}, \bibinfo {author}
  {\bibfnamefont {Jin-Shi}\ \bibnamefont {Xu}}, \bibinfo {author}
  {\bibfnamefont {Chuan-Feng}\ \bibnamefont {Li}}, \ and\ \bibinfo {author}
  {\bibfnamefont {Guang-Can}\ \bibnamefont {Guo}},\ }\bibfield  {title}
  {\enquote {\bibinfo {title} {Observing half-integer topological winding
  numbers in non-hermitian synthetic lattices},}\ }\href@noop {} {\bibfield
  {journal} {\bibinfo  {journal} {Light: Science \& Applications}\ }\textbf
  {\bibinfo {volume} {14}},\ \bibinfo {pages} {225} (\bibinfo {year}
  {2025})}\BibitemShut {NoStop}%
\bibitem [{\citenamefont {Liang}\ and\ \citenamefont
  {Li}(2025)}]{liang2025topological}%
  \BibitemOpen
  \bibfield  {author} {\bibinfo {author} {\bibfnamefont {Gan}\ \bibnamefont
  {Liang}}\ and\ \bibinfo {author} {\bibfnamefont {Linhu}\ \bibnamefont {Li}},\
  }\bibfield  {title} {\enquote {\bibinfo {title} {Topological and fractal
  defect states in non-hermitian lattices},}\ }\href@noop {} {\bibfield
  {journal} {\bibinfo  {journal} {Physical Review B}\ }\textbf {\bibinfo
  {volume} {112}},\ \bibinfo {pages} {045115} (\bibinfo {year}
  {2025})}\BibitemShut {NoStop}%
\bibitem [{\citenamefont {Yao}\ and\ \citenamefont {Wang}(2018)}]{yao2018edge}%
  \BibitemOpen
  \bibfield  {author} {\bibinfo {author} {\bibfnamefont {Shunyu}\ \bibnamefont
  {Yao}}\ and\ \bibinfo {author} {\bibfnamefont {Zhong}\ \bibnamefont {Wang}},\
  }\bibfield  {title} {\enquote {\bibinfo {title} {Edge states and topological
  invariants of non-hermitian systems},}\ }\href@noop {} {\bibfield  {journal}
  {\bibinfo  {journal} {Physical review letters}\ }\textbf {\bibinfo {volume}
  {121}},\ \bibinfo {pages} {086803} (\bibinfo {year} {2018})}\BibitemShut
  {NoStop}%
\bibitem [{\citenamefont {Yao}\ \emph {et~al.}(2018)\citenamefont {Yao},
  \citenamefont {Song},\ and\ \citenamefont {Wang}}]{yao2018non}%
  \BibitemOpen
  \bibfield  {author} {\bibinfo {author} {\bibfnamefont {Shunyu}\ \bibnamefont
  {Yao}}, \bibinfo {author} {\bibfnamefont {Fei}\ \bibnamefont {Song}}, \ and\
  \bibinfo {author} {\bibfnamefont {Zhong}\ \bibnamefont {Wang}},\ }\bibfield
  {title} {\enquote {\bibinfo {title} {Non-hermitian chern bands},}\
  }\href@noop {} {\bibfield  {journal} {\bibinfo  {journal} {Physical review
  letters}\ }\textbf {\bibinfo {volume} {121}},\ \bibinfo {pages} {136802}
  (\bibinfo {year} {2018})}\BibitemShut {NoStop}%
\bibitem [{\citenamefont {Martinez~Alvarez}\ \emph {et~al.}(2018)\citenamefont
  {Martinez~Alvarez}, \citenamefont {Barrios~Vargas},\ and\ \citenamefont
  {Foa~Torres}}]{martinez2018non}%
  \BibitemOpen
  \bibfield  {author} {\bibinfo {author} {\bibfnamefont {VM}~\bibnamefont
  {Martinez~Alvarez}}, \bibinfo {author} {\bibfnamefont {JE}~\bibnamefont
  {Barrios~Vargas}}, \ and\ \bibinfo {author} {\bibfnamefont {LEF}\
  \bibnamefont {Foa~Torres}},\ }\bibfield  {title} {\enquote {\bibinfo {title}
  {Non-hermitian robust edge states in one dimension: Anomalous localization
  and eigenspace condensation at exceptional points},}\ }\href@noop {}
  {\bibfield  {journal} {\bibinfo  {journal} {Physical review B}\ }\textbf
  {\bibinfo {volume} {97}},\ \bibinfo {pages} {121401} (\bibinfo {year}
  {2018})}\BibitemShut {NoStop}%
\bibitem [{\citenamefont {Jin}\ and\ \citenamefont {Song}(2019)}]{jin2019bulk}%
  \BibitemOpen
  \bibfield  {author} {\bibinfo {author} {\bibfnamefont {L.}~\bibnamefont
  {Jin}}\ and\ \bibinfo {author} {\bibfnamefont {Z.}~\bibnamefont {Song}},\
  }\bibfield  {title} {\enquote {\bibinfo {title} {Bulk-boundary correspondence
  in a non-hermitian system in one dimension with chiral inversion symmetry},}\
  }\href {\doibase 10.1103/PhysRevB.99.081103} {\bibfield  {journal} {\bibinfo
  {journal} {Phys. Rev. B}\ }\textbf {\bibinfo {volume} {99}},\ \bibinfo
  {pages} {081103} (\bibinfo {year} {2019})}\BibitemShut {NoStop}%
\bibitem [{\citenamefont {Lee}\ and\ \citenamefont
  {Thomale}(2019)}]{lee2019anatomy}%
  \BibitemOpen
  \bibfield  {author} {\bibinfo {author} {\bibfnamefont {Ching~Hua}\
  \bibnamefont {Lee}}\ and\ \bibinfo {author} {\bibfnamefont {Ronny}\
  \bibnamefont {Thomale}},\ }\bibfield  {title} {\enquote {\bibinfo {title}
  {Anatomy of skin modes and topology in non-hermitian systems},}\ }\href@noop
  {} {\bibfield  {journal} {\bibinfo  {journal} {Physical Review B}\ }\textbf
  {\bibinfo {volume} {99}},\ \bibinfo {pages} {201103} (\bibinfo {year}
  {2019})}\BibitemShut {NoStop}%
\bibitem [{\citenamefont {Okuma}\ and\ \citenamefont
  {Sato}(2023)}]{okuma2023non}%
  \BibitemOpen
  \bibfield  {author} {\bibinfo {author} {\bibfnamefont {Nobuyuki}\
  \bibnamefont {Okuma}}\ and\ \bibinfo {author} {\bibfnamefont {Masatoshi}\
  \bibnamefont {Sato}},\ }\bibfield  {title} {\enquote {\bibinfo {title}
  {Non-hermitian topological phenomena: A review},}\ }\href@noop {} {\bibfield
  {journal} {\bibinfo  {journal} {Annual Review of Condensed Matter Physics}\
  }\textbf {\bibinfo {volume} {14}},\ \bibinfo {pages} {83--107} (\bibinfo
  {year} {2023})}\BibitemShut {NoStop}%
\bibitem [{\citenamefont {Ding}\ \emph {et~al.}(2022)\citenamefont {Ding},
  \citenamefont {Fang},\ and\ \citenamefont {Ma}}]{ding2022non}%
  \BibitemOpen
  \bibfield  {author} {\bibinfo {author} {\bibfnamefont {Kun}\ \bibnamefont
  {Ding}}, \bibinfo {author} {\bibfnamefont {Chen}\ \bibnamefont {Fang}}, \
  and\ \bibinfo {author} {\bibfnamefont {Guancong}\ \bibnamefont {Ma}},\
  }\bibfield  {title} {\enquote {\bibinfo {title} {Non-hermitian topology and
  exceptional-point geometries},}\ }\href@noop {} {\bibfield  {journal}
  {\bibinfo  {journal} {Nature Reviews Physics}\ }\textbf {\bibinfo {volume}
  {4}},\ \bibinfo {pages} {745--760} (\bibinfo {year} {2022})}\BibitemShut
  {NoStop}%
\bibitem [{\citenamefont {Zhang}\ \emph {et~al.}(2022)\citenamefont {Zhang},
  \citenamefont {Zhang}, \citenamefont {Lu},\ and\ \citenamefont
  {Chen}}]{zhang2022review}%
  \BibitemOpen
  \bibfield  {author} {\bibinfo {author} {\bibfnamefont {Xiujuan}\ \bibnamefont
  {Zhang}}, \bibinfo {author} {\bibfnamefont {Tian}\ \bibnamefont {Zhang}},
  \bibinfo {author} {\bibfnamefont {Ming-Hui}\ \bibnamefont {Lu}}, \ and\
  \bibinfo {author} {\bibfnamefont {Yan-Feng}\ \bibnamefont {Chen}},\
  }\bibfield  {title} {\enquote {\bibinfo {title} {A review on non-hermitian
  skin effect},}\ }\href@noop {} {\bibfield  {journal} {\bibinfo  {journal}
  {Advances in Physics: X}\ }\textbf {\bibinfo {volume} {7}},\ \bibinfo {pages}
  {2109431} (\bibinfo {year} {2022})}\BibitemShut {NoStop}%
\bibitem [{\citenamefont {Lin}\ \emph {et~al.}(2023)\citenamefont {Lin},
  \citenamefont {Tai}, \citenamefont {Li},\ and\ \citenamefont
  {Lee}}]{lin2023topological}%
  \BibitemOpen
  \bibfield  {author} {\bibinfo {author} {\bibfnamefont {Rijia}\ \bibnamefont
  {Lin}}, \bibinfo {author} {\bibfnamefont {Tommy}\ \bibnamefont {Tai}},
  \bibinfo {author} {\bibfnamefont {Linhu}\ \bibnamefont {Li}}, \ and\ \bibinfo
  {author} {\bibfnamefont {Ching~Hua}\ \bibnamefont {Lee}},\ }\bibfield
  {title} {\enquote {\bibinfo {title} {Topological non-hermitian skin
  effect},}\ }\href@noop {} {\bibfield  {journal} {\bibinfo  {journal}
  {Frontiers of Physics}\ }\textbf {\bibinfo {volume} {18}},\ \bibinfo {pages}
  {53605} (\bibinfo {year} {2023})}\BibitemShut {NoStop}%
\bibitem [{\citenamefont {Li}\ \emph {et~al.}(2021)\citenamefont {Li},
  \citenamefont {Mu}, \citenamefont {Lee},\ and\ \citenamefont
  {Gong}}]{li2021quantized}%
  \BibitemOpen
  \bibfield  {author} {\bibinfo {author} {\bibfnamefont {Linhu}\ \bibnamefont
  {Li}}, \bibinfo {author} {\bibfnamefont {Sen}\ \bibnamefont {Mu}}, \bibinfo
  {author} {\bibfnamefont {Ching~Hua}\ \bibnamefont {Lee}}, \ and\ \bibinfo
  {author} {\bibfnamefont {Jiangbin}\ \bibnamefont {Gong}},\ }\bibfield
  {title} {\enquote {\bibinfo {title} {Quantized classical response from
  spectral winding topology},}\ }\href@noop {} {\bibfield  {journal} {\bibinfo
  {journal} {Nature communications}\ }\textbf {\bibinfo {volume} {12}},\
  \bibinfo {pages} {5294} (\bibinfo {year} {2021})}\BibitemShut {NoStop}%
\bibitem [{\citenamefont {Liu}\ \emph {et~al.}(2021)\citenamefont {Liu},
  \citenamefont {Zeng}, \citenamefont {Li},\ and\ \citenamefont
  {Chen}}]{liu2021exact}%
  \BibitemOpen
  \bibfield  {author} {\bibinfo {author} {\bibfnamefont {Yanxia}\ \bibnamefont
  {Liu}}, \bibinfo {author} {\bibfnamefont {Yumeng}\ \bibnamefont {Zeng}},
  \bibinfo {author} {\bibfnamefont {Linhu}\ \bibnamefont {Li}}, \ and\ \bibinfo
  {author} {\bibfnamefont {Shu}\ \bibnamefont {Chen}},\ }\bibfield  {title}
  {\enquote {\bibinfo {title} {Exact solution of the single impurity problem in
  nonreciprocal lattices: Impurity-induced size-dependent non-hermitian skin
  effect},}\ }\href@noop {} {\bibfield  {journal} {\bibinfo  {journal}
  {Physical Review B}\ }\textbf {\bibinfo {volume} {104}},\ \bibinfo {pages}
  {085401} (\bibinfo {year} {2021})}\BibitemShut {NoStop}%
\bibitem [{\citenamefont {Liang}\ \emph {et~al.}(2022)\citenamefont {Liang},
  \citenamefont {Mu}, \citenamefont {Gong},\ and\ \citenamefont
  {Li}}]{liang2022anomalous}%
  \BibitemOpen
  \bibfield  {author} {\bibinfo {author} {\bibfnamefont {Hui-Qiang}\
  \bibnamefont {Liang}}, \bibinfo {author} {\bibfnamefont {Sen}\ \bibnamefont
  {Mu}}, \bibinfo {author} {\bibfnamefont {Jiangbin}\ \bibnamefont {Gong}}, \
  and\ \bibinfo {author} {\bibfnamefont {Linhu}\ \bibnamefont {Li}},\
  }\bibfield  {title} {\enquote {\bibinfo {title} {Anomalous hybridization of
  spectral winding topology in quantized steady-state responses},}\ }\href@noop
  {} {\bibfield  {journal} {\bibinfo  {journal} {Physical Review B}\ }\textbf
  {\bibinfo {volume} {105}},\ \bibinfo {pages} {L241402} (\bibinfo {year}
  {2022})}\BibitemShut {NoStop}%
\bibitem [{\citenamefont {Ou}\ \emph {et~al.}(2023)\citenamefont {Ou},
  \citenamefont {Wang},\ and\ \citenamefont {Li}}]{ou2023non}%
  \BibitemOpen
  \bibfield  {author} {\bibinfo {author} {\bibfnamefont {Zuxuan}\ \bibnamefont
  {Ou}}, \bibinfo {author} {\bibfnamefont {Yucheng}\ \bibnamefont {Wang}}, \
  and\ \bibinfo {author} {\bibfnamefont {Linhu}\ \bibnamefont {Li}},\
  }\bibfield  {title} {\enquote {\bibinfo {title} {Non-hermitian boundary
  spectral winding},}\ }\href@noop {} {\bibfield  {journal} {\bibinfo
  {journal} {Physical Review B}\ }\textbf {\bibinfo {volume} {107}},\ \bibinfo
  {pages} {L161404} (\bibinfo {year} {2023})}\BibitemShut {NoStop}%
\bibitem [{\citenamefont {Economou}(2006)}]{economou2006green}%
  \BibitemOpen
  \bibfield  {author} {\bibinfo {author} {\bibfnamefont {Eleftherios~N}\
  \bibnamefont {Economou}},\ }\href@noop {} {\emph {\bibinfo {title} {Green’s
  functions in quantum physics}}}\ (\bibinfo  {publisher} {Springer},\ \bibinfo
  {year} {2006})\BibitemShut {NoStop}%
\bibitem [{\citenamefont {Leonforte}\ \emph {et~al.}(2021)\citenamefont
  {Leonforte}, \citenamefont {Carollo},\ and\ \citenamefont
  {Ciccarello}}]{leonforte2021vacancy}%
  \BibitemOpen
  \bibfield  {author} {\bibinfo {author} {\bibfnamefont {Luca}\ \bibnamefont
  {Leonforte}}, \bibinfo {author} {\bibfnamefont {Angelo}\ \bibnamefont
  {Carollo}}, \ and\ \bibinfo {author} {\bibfnamefont {Francesco}\ \bibnamefont
  {Ciccarello}},\ }\bibfield  {title} {\enquote {\bibinfo {title} {Vacancy-like
  dressed states in topological waveguide qed},}\ }\href@noop {} {\bibfield
  {journal} {\bibinfo  {journal} {Physical Review Letters}\ }\textbf {\bibinfo
  {volume} {126}},\ \bibinfo {pages} {063601} (\bibinfo {year}
  {2021})}\BibitemShut {NoStop}%
\bibitem [{\citenamefont {Lombardo}\ \emph {et~al.}(2014)\citenamefont
  {Lombardo}, \citenamefont {Ciccarello},\ and\ \citenamefont
  {Palma}}]{lombardo2014photon}%
  \BibitemOpen
  \bibfield  {author} {\bibinfo {author} {\bibfnamefont {Federico}\
  \bibnamefont {Lombardo}}, \bibinfo {author} {\bibfnamefont {Francesco}\
  \bibnamefont {Ciccarello}}, \ and\ \bibinfo {author} {\bibfnamefont
  {G~Massimo}\ \bibnamefont {Palma}},\ }\bibfield  {title} {\enquote {\bibinfo
  {title} {Photon localization versus population trapping in a coupled-cavity
  array},}\ }\href@noop {} {\bibfield  {journal} {\bibinfo  {journal} {Physical
  Review A}\ }\textbf {\bibinfo {volume} {89}},\ \bibinfo {pages} {053826}
  (\bibinfo {year} {2014})}\BibitemShut {NoStop}%
\bibitem [{\citenamefont {Roccati}(2021)}]{roccati2021non}%
  \BibitemOpen
  \bibfield  {author} {\bibinfo {author} {\bibfnamefont {Federico}\
  \bibnamefont {Roccati}},\ }\bibfield  {title} {\enquote {\bibinfo {title}
  {Non-hermitian skin effect as an impurity problem},}\ }\href@noop {}
  {\bibfield  {journal} {\bibinfo  {journal} {Physical Review A}\ }\textbf
  {\bibinfo {volume} {104}},\ \bibinfo {pages} {022215} (\bibinfo {year}
  {2021})}\BibitemShut {NoStop}%
\bibitem [{Sup()}]{SuppMat}%
  \BibitemOpen
  \href@noop {} {\bibinfo  {journal} {Supplemental Materials which contain
  Refs. \cite{kunst2018biorthogonal,brody2013biorthogonal} in addition to those
  also appear in the main text.}\ }\BibitemShut {NoStop}%
\bibitem [{\citenamefont {Wanjura}\ \emph {et~al.}(2020)\citenamefont
  {Wanjura}, \citenamefont {Brunelli},\ and\ \citenamefont
  {Nunnenkamp}}]{wanjura2020topological}%
  \BibitemOpen
\bibfield  {journal} {  }\bibfield  {author} {\bibinfo {author} {\bibfnamefont
  {Clara~C}\ \bibnamefont {Wanjura}}, \bibinfo {author} {\bibfnamefont
  {Matteo}\ \bibnamefont {Brunelli}}, \ and\ \bibinfo {author} {\bibfnamefont
  {Andreas}\ \bibnamefont {Nunnenkamp}},\ }\bibfield  {title} {\enquote
  {\bibinfo {title} {Topological framework for directional amplification in
  driven-dissipative cavity arrays},}\ }\href@noop {} {\bibfield  {journal}
  {\bibinfo  {journal} {Nature communications}\ }\textbf {\bibinfo {volume}
  {11}},\ \bibinfo {pages} {3149} (\bibinfo {year} {2020})}\BibitemShut
  {NoStop}%
\bibitem [{\citenamefont {Xue}\ \emph {et~al.}(2021)\citenamefont {Xue},
  \citenamefont {Li}, \citenamefont {Hu}, \citenamefont {Song},\ and\
  \citenamefont {Wang}}]{xue2021simple}%
  \BibitemOpen
  \bibfield  {author} {\bibinfo {author} {\bibfnamefont {Wen-Tan}\ \bibnamefont
  {Xue}}, \bibinfo {author} {\bibfnamefont {Ming-Rui}\ \bibnamefont {Li}},
  \bibinfo {author} {\bibfnamefont {Yu-Min}\ \bibnamefont {Hu}}, \bibinfo
  {author} {\bibfnamefont {Fei}\ \bibnamefont {Song}}, \ and\ \bibinfo {author}
  {\bibfnamefont {Zhong}\ \bibnamefont {Wang}},\ }\bibfield  {title} {\enquote
  {\bibinfo {title} {Simple formulas of directional amplification from
  non-bloch band theory},}\ }\href@noop {} {\bibfield  {journal} {\bibinfo
  {journal} {Physical Review B}\ }\textbf {\bibinfo {volume} {103}},\ \bibinfo
  {pages} {L241408} (\bibinfo {year} {2021})}\BibitemShut {NoStop}%
\bibitem [{\citenamefont {Wanjura}\ \emph {et~al.}(2021)\citenamefont
  {Wanjura}, \citenamefont {Brunelli},\ and\ \citenamefont
  {Nunnenkamp}}]{wanjura2021correspondence}%
  \BibitemOpen
  \bibfield  {author} {\bibinfo {author} {\bibfnamefont {Clara~C}\ \bibnamefont
  {Wanjura}}, \bibinfo {author} {\bibfnamefont {Matteo}\ \bibnamefont
  {Brunelli}}, \ and\ \bibinfo {author} {\bibfnamefont {Andreas}\ \bibnamefont
  {Nunnenkamp}},\ }\bibfield  {title} {\enquote {\bibinfo {title}
  {Correspondence between non-hermitian topology and directional amplification
  in the presence of disorder},}\ }\href@noop {} {\bibfield  {journal}
  {\bibinfo  {journal} {Physical Review Letters}\ }\textbf {\bibinfo {volume}
  {127}},\ \bibinfo {pages} {213601} (\bibinfo {year} {2021})}\BibitemShut
  {NoStop}%
\bibitem [{\citenamefont {Lee}\ \emph {et~al.}(2019)\citenamefont {Lee},
  \citenamefont {Ahn}, \citenamefont {Zhou},\ and\ \citenamefont
  {Vishwanath}}]{lee2019topological}%
  \BibitemOpen
  \bibfield  {author} {\bibinfo {author} {\bibfnamefont {Jong~Yeon}\
  \bibnamefont {Lee}}, \bibinfo {author} {\bibfnamefont {Junyeong}\
  \bibnamefont {Ahn}}, \bibinfo {author} {\bibfnamefont {Hengyun}\ \bibnamefont
  {Zhou}}, \ and\ \bibinfo {author} {\bibfnamefont {Ashvin}\ \bibnamefont
  {Vishwanath}},\ }\bibfield  {title} {\enquote {\bibinfo {title} {Topological
  correspondence between hermitian and non-hermitian systems: Anomalous
  dynamics},}\ }\href@noop {} {\bibfield  {journal} {\bibinfo  {journal}
  {Physical review letters}\ }\textbf {\bibinfo {volume} {123}},\ \bibinfo
  {pages} {206404} (\bibinfo {year} {2019})}\BibitemShut {NoStop}%
\bibitem [{\citenamefont {Schnyder}\ \emph {et~al.}(2008)\citenamefont
  {Schnyder}, \citenamefont {Ryu}, \citenamefont {Furusaki},\ and\
  \citenamefont {Ludwig}}]{schnyder2008classification}%
  \BibitemOpen
  \bibfield  {author} {\bibinfo {author} {\bibfnamefont {Andreas~P}\
  \bibnamefont {Schnyder}}, \bibinfo {author} {\bibfnamefont {Shinsei}\
  \bibnamefont {Ryu}}, \bibinfo {author} {\bibfnamefont {Akira}\ \bibnamefont
  {Furusaki}}, \ and\ \bibinfo {author} {\bibfnamefont {Andreas~WW}\
  \bibnamefont {Ludwig}},\ }\bibfield  {title} {\enquote {\bibinfo {title}
  {Classification of topological insulators and superconductors in three
  spatial dimensions},}\ }\href@noop {} {\bibfield  {journal} {\bibinfo
  {journal} {Physical Review B—Condensed Matter and Materials Physics}\
  }\textbf {\bibinfo {volume} {78}},\ \bibinfo {pages} {195125} (\bibinfo
  {year} {2008})}\BibitemShut {NoStop}%
\bibitem [{\citenamefont {Ryu}\ \emph {et~al.}(2010)\citenamefont {Ryu},
  \citenamefont {Schnyder}, \citenamefont {Furusaki},\ and\ \citenamefont
  {Ludwig}}]{ryu2010topological}%
  \BibitemOpen
  \bibfield  {author} {\bibinfo {author} {\bibfnamefont {Shinsei}\ \bibnamefont
  {Ryu}}, \bibinfo {author} {\bibfnamefont {Andreas~P}\ \bibnamefont
  {Schnyder}}, \bibinfo {author} {\bibfnamefont {Akira}\ \bibnamefont
  {Furusaki}}, \ and\ \bibinfo {author} {\bibfnamefont {Andreas~WW}\
  \bibnamefont {Ludwig}},\ }\bibfield  {title} {\enquote {\bibinfo {title}
  {Topological insulators and superconductors: tenfold way and dimensional
  hierarchy},}\ }\href@noop {} {\bibfield  {journal} {\bibinfo  {journal} {New
  Journal of Physics}\ }\textbf {\bibinfo {volume} {12}},\ \bibinfo {pages}
  {065010} (\bibinfo {year} {2010})}\BibitemShut {NoStop}%
\bibitem [{\citenamefont {Kunst}\ \emph {et~al.}(2018)\citenamefont {Kunst},
  \citenamefont {Edvardsson}, \citenamefont {Budich},\ and\ \citenamefont
  {Bergholtz}}]{kunst2018biorthogonal}%
  \BibitemOpen
  \bibfield  {author} {\bibinfo {author} {\bibfnamefont {Flore~K}\ \bibnamefont
  {Kunst}}, \bibinfo {author} {\bibfnamefont {Elisabet}\ \bibnamefont
  {Edvardsson}}, \bibinfo {author} {\bibfnamefont {Jan~Carl}\ \bibnamefont
  {Budich}}, \ and\ \bibinfo {author} {\bibfnamefont {Emil~J}\ \bibnamefont
  {Bergholtz}},\ }\bibfield  {title} {\enquote {\bibinfo {title} {Biorthogonal
  bulk-boundary correspondence in non-hermitian systems},}\ }\href@noop {}
  {\bibfield  {journal} {\bibinfo  {journal} {Physical review letters}\
  }\textbf {\bibinfo {volume} {121}},\ \bibinfo {pages} {026808} (\bibinfo
  {year} {2018})}\BibitemShut {NoStop}%
\bibitem [{\citenamefont {Brody}(2013)}]{brody2013biorthogonal}%
  \BibitemOpen
  \bibfield  {author} {\bibinfo {author} {\bibfnamefont {Dorje~C}\ \bibnamefont
  {Brody}},\ }\bibfield  {title} {\enquote {\bibinfo {title} {Biorthogonal
  quantum mechanics},}\ }\href@noop {} {\bibfield  {journal} {\bibinfo
  {journal} {Journal of Physics A: Mathematical and Theoretical}\ }\textbf
  {\bibinfo {volume} {47}},\ \bibinfo {pages} {035305} (\bibinfo {year}
  {2013})}\BibitemShut {NoStop}%
\end{thebibliography}%

\end{document}